\documentclass[aps,prl,reprint,superscriptaddress]{revtex4-2}

\usepackage[english]{babel}
\usepackage{graphics}
\usepackage{graphicx}
\usepackage{amsfonts}
\usepackage{amsmath}
\usepackage{amssymb}
\usepackage{color}
\usepackage{todonotes}

\definecolor{darkblue}{rgb}{0.0,0.0,0.4}
\definecolor{darkgreen}{rgb}{0.0,0.4,0.0}
\definecolor{darkred}{rgb}{0.6,0.0,0.0}
\usepackage[bookmarksopen]{hyperref}
\hypersetup{colorlinks,linkcolor=darkblue,citecolor=darkblue,urlcolor=darkblue}

\newcommand{\gs}{{$X^{2}\Sigma^{+}$}}
\newcommand{\exs}{{$A{}^{2}\Pi_{1/2}$}}

\newcommand{\baff}[1]{${}^{#1}\mathrm{BaF}$}
\newcommand\thefontsize{The current font size is: \f@size pt}

\begin{document}

\title{Isotopologue-selective laser cooling of molecules}

\author{Felix Kogel}
\thanks{These two authors contributed equally}
\affiliation{5. Physikalisches  Institut  and  Center  for  Integrated  Quantum  Science  and  Technology, Universit\"at  Stuttgart,  Pfaffenwaldring  57,  70569  Stuttgart,  Germany}

\author{Tatsam Garg}
\thanks{These two authors contributed equally}
\affiliation{5. Physikalisches  Institut  and  Center  for  Integrated  Quantum  Science  and  Technology, Universit\"at  Stuttgart,  Pfaffenwaldring  57,  70569  Stuttgart,  Germany}

\author{Marian Rockenh\"auser} 
\affiliation{5. Physikalisches  Institut  and  Center  for  Integrated  Quantum  Science  and  Technology, Universit\"at  Stuttgart,  Pfaffenwaldring  57,  70569  Stuttgart,  Germany}

\affiliation{Vienna Center for Quantum Science and Technology, Atominstitut, TU Wien,  Stadionallee 2,  A-1020 Vienna,  Austria}

\author{Sebasti\'an A. Morales-Ram\'irez}
\affiliation{5. Physikalisches  Institut  and  Center  for  Integrated  Quantum  Science  and  Technology, Universit\"at  Stuttgart,  Pfaffenwaldring  57,  70569  Stuttgart,  Germany}

\author{Tim Langen}
\email{tim.langen@tuwien.ac.at}

\affiliation{5. Physikalisches  Institut  and  Center  for  Integrated  Quantum  Science  and  Technology, Universit\"at  Stuttgart,  Pfaffenwaldring  57,  70569  Stuttgart,  Germany}

\affiliation{Vienna Center for Quantum Science and Technology, Atominstitut, TU Wien,  Stadionallee 2,  A-1020 Vienna,  Austria}

\begin{abstract}
Direct laser cooling of molecules has made significant progress in recent years. 
However, the selective cooling and manipulation of molecules based on their isotopic composition, which is ubiquitous in atomic laser cooling, has not yet been achieved. Here, we demonstrate such isotopologue-selective laser cooling of molecules, using barium monofluoride (BaF) as an example. The manipulation of the rare and previously uncooled \baff{136} is achieved within a molecular beam containing several isotopologues of significantly higher natural abundance. Our results enable intense molecular beams and high fidelity detection of select low-abundance isotopologues or isotopic mixtures. Such beams are a first step towards isotopologue-selective molecular trapping and will be useful for applications in trace gas analysis, cold chemistry and precision tests of fundamental symmetries. 
\end{abstract}

\maketitle

\section{Introduction}
Isotope-selective laser cooling is a standard technique in atomic physics. It is facilitated by the simple spectra and isotope shifts in atoms and plays a key role for experiments with isotopic atomic mixtures~\cite{Poli2005,Papp2008,Mickelson2010}, the cooling of atoms with specific quantum statistics~\cite{Schreck2001,Truscott2001}, and isotope-specific collision experiments~\cite{Kato2017,Tanzi2018,Durastante2020}. Moreover, the high specificity of atomic laser cooling and trapping is also widely used for isotope separation and collection~\cite{Bernhardt1974}, facilitating the study of rare isotopes~\cite{Lu1994,Gwinner1994,Simsarian1996,Lu1997,Guest2007}, trace gas analysis~\cite{Chen1999,Hoekstra2005,Welte2010}, and precision measurements~\cite{Scielzo2004,Gorelov2005,Mueller2007,Wilschut2010,Martoff2021}. Notably, the technique can accumulate significant numbers of atoms even from the extremely low output of sources based on nuclear decay or accelerators~\cite{Lu1994,Gwinner1994}. 

Recently, important progress has been made in transferring such laser cooling techniques from atoms to molecules~\cite{Shuman2010,Fitch2021,McCarron2018a,Langen2023}. As a result, both diatomic and polyatomic species of ever-increasing complexity have been cooled,  trapped~\cite{Barry2014,Collopy2018,Vilas2021}, and loaded into magnetic traps~\cite{McCarron2018,Williams2018}, optical lattices~\cite{Wu2021} and tweezers~\cite{Anderegg2019,Vilas2023}. This has facilitated studies of collisions down to the single molecule and single quantum state level~\cite{Cheuk2020}. However, the selective application of laser cooling to mixtures of different, and potentially rare isotopologues of the same molecular species has so far remained elusive. 

Here, we use a beam of barium monofluoride (BaF) molecules that contains several naturally occurring stable isotopologues to demonstrate selective manipulation and laser cooling of its low-abundance \baff{136} isotopologue. Our work has immediate implications for precision measurements using BaF molecules~\cite{Skripnikov2014,Lackenby2018,ArrowsmithKron2023,Tiberi2024}, where isotopic variations can help to disentangle competing fundamental processes in searches for physics beyond the Standard Model~\cite{Demille2008,Altuntas2018,Antypas2019,Hao2020}. Furthermore, by generalizing the principles outlined in the following, our work paves the way for isotopologue-selective laser trapping, through which long-lived trapped samples and mixtures of many molecular species can become available. This opens up a new approach to the study of cold collisions and isotope exchange chemistry~\cite{Tomza2015,Fleurat2003,Brenninkmeijer2003,Fleischer2006,Visser2009,Zhdanovich2012}, trace gas analysis~\cite{Griffith2018} and applications in astrochemistry~\cite{Kaminsky2018}.

\begin{figure}[tb]
    \centering
    \includegraphics[width=0.99\columnwidth]{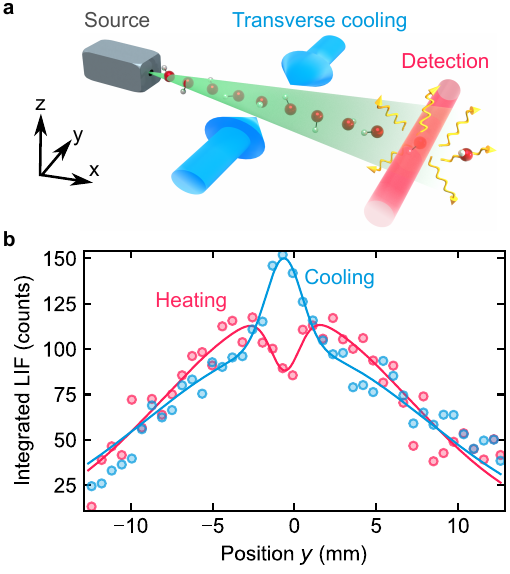}
    \caption{(a) Experimental setup. Molecules are created in a cryogenic buffer gas source, pass a laser cooling region where they interact with transverse laser beams, and are subsequently detected by laser-induced fluorescence (LIF). (b)~Transverse molecular distribution of the rare \baff{136} isotopologue recorded via LIF, revealing the typical signature of Sisyphus cooling, where molecules accumulate in the coldest part of the beam for blue-detuned cooling light. In contrast to this, Sisyphus heating appears for red-detuned cooling light. Datapoints are averaged over 200 experimental realizations.\vspace{-20pt}}
    \label{fig:cooling}
\end{figure} 

\section{Experiment}
Our experimental setup to create and manipulate a BaF molecular beam is summarized in Fig.~\ref{fig:cooling}a. It is based on a standard cryogenic buffer gas source and has been described in detail elsewhere~\cite{Albrecht2020,Rockenhaeuser2023}. In short, we ablate solid $\mathrm{BaF}_2$ targets containing a large variety of naturally occuring barium isotopes into cryogenic helium gas and collimate the resulting BaF molecular beam. The beam expands into a room-temperature vacuum chamber with a transverse temperature of approximately $100\,$mK. In this chamber, it can be manipulated transversally using laser beams. Subsequently, the transverse distribution of the molecules is characterized by laser-induced fluorescence that is recorded on an EMCCD camera. 

Example results of the cooling process are shown in Fig.~\ref{fig:cooling}b, where magnetically-assisted Sisyphus laser cooling forces have been used to efficiently cool or heat the transverse motion of the \baff{136} molecules~\cite{Fitch2021,Rockenhaeuser2024}. In the cooling configuration a large number of molecules are accumulated in the coldest part of the molecular beam, which could readily be used for further manipulation and experiments. The upper temperature limit resolvable for these molecules in the experiment is on the order of $2.5\,$mK, with the actual temperature expected to be significantly lower~\cite{Rockenhaeuser2024}. Cooling does not interfere with the other isotopologues present in the same molecular beam, in particular the order of magnitude more abundant \baff{138}. This is achieved while maintaining full quantum state control over all isotopologues, highlighting the specificity and versatility of the technique. 

\section{Laser cooling and detection scheme}

The BaF molecules used in our experiment are formed by a predominantly ionic bond between a barium and a fluorine atom. 
While there is only a single stable fluorine isotope ($^{19}$F), barium features a wide variety of stable isotopes. This results in five BaF isotopologues, \baff{134} through \baff{138}, with natural abundances above $1\%$. By far the most common of these is the bosonic \baff{138} with an abundance of $71.7\%$. This isotopologue has been extensively studied in experiments~\cite{Bu2017,Aggarwal2018,Albrecht2020,Cournol2018,Hao2019,Courageux2022,Denis2022,Li2023,Rockenhaeuser2023} and recently also laser cooled \cite{Rockenhaeuser2024}. Significantly rarer, but with a similar level structure, is \baff{136}, which has an abundance of  $7.85\%$. This combination of known level structure and comparatively low, but not vanishing, abundance makes \baff{136} an attractive testing ground for isotopologue-selective laser cooling. 

\begin{figure}[tb]
    \centering
    \includegraphics[width=\columnwidth]{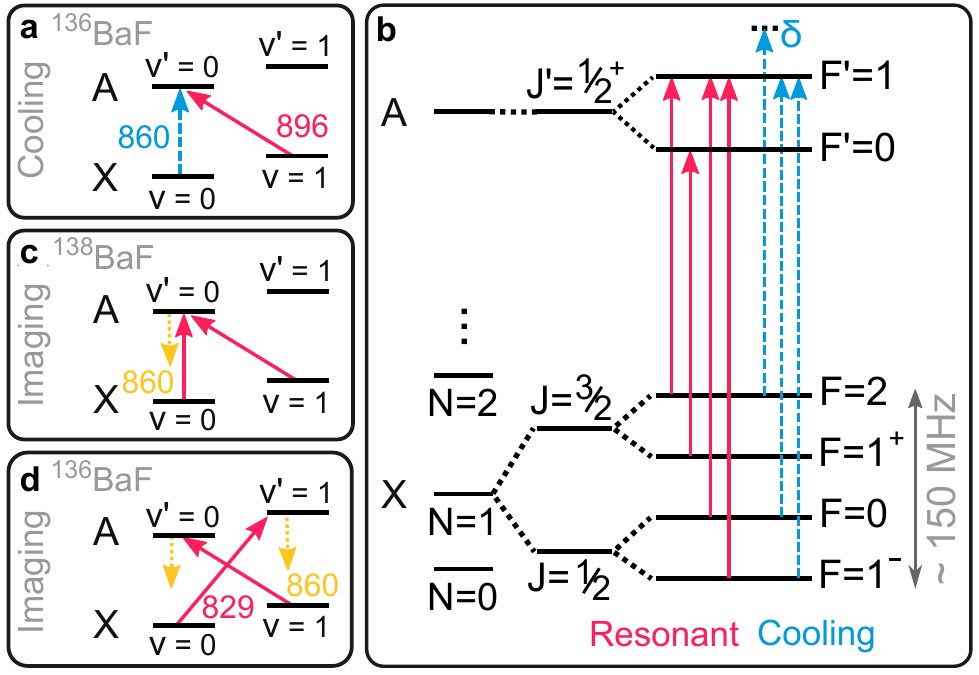}
    \caption{Level structure of BaF. (a) In addition to the main cooling laser, which addresses the X-A electronic transition without a change of vibrational quantum number (dashed blue arrow, $\nu=0\rightarrow\nu'=0$), laser cooling of molecules requires vibrational repumpers (solid red arrow, $\nu=1\rightarrow\nu'=0$) to limit vibrational branching. (b) Furthermore, to realize a closed optical cycle, rotational branching needs to be suppressed by driving transitions between the lowest odd parity rotational state ($N=1$) in the ground state and the lowest even parity state ($J=1/2^+$) in the excited state. Further hyperfine substructure with splittings on the order of $150\,$MHz is addressed using laser sidebands. These sidebands can either be near-resonant with all transitions for repumping and detection (red solid arrows), or address only some transitions and include a detuning $\delta$ to realize cooling forces (blue dashed arrows). (c) Detection for high-abundance \baff{138} is implemented by combining resonant cooler and repumper, with fluorescence (dotted arrow pointing down) emitted on the cooling transition, and (d) for the rare \baff{136} by combining a resonant depumper (solid arrow, $\nu=0\rightarrow\nu'=1$) and repumper in a Raman cycling scheme, that yields spectrally separated fluorescence. The latter minimizes the effect of stray light and facilitates imaging of rare isotopologues with high signal-to-noise ratio. Numbers denote the wavelengths of the respective transitions for BaF in nanometers. Energy differences are not to scale.}
    \label{fig:cyclingscheme}
\end{figure}

To realize laser cooling of BaF molecules, we drive electronic transitions between their \gs\, ground state and \exs\, excited state. These transitions have a wavelength of around $860\,$nm, where ample power is available from simple diode laser systems. However, to make the transitions quasi-closed, and thus suitable for the scattering of a sufficiently large number of photons for laser cooling and high-fidelity detection, both vibrational and rotational branching, as well as potential hyperfine structure dark states must be addressed. The relevant level structure for this is summarized in Fig.~\ref{fig:cyclingscheme}. 

\begin{figure*}[tb]
    \centering
    \includegraphics[width=0.93\textwidth]{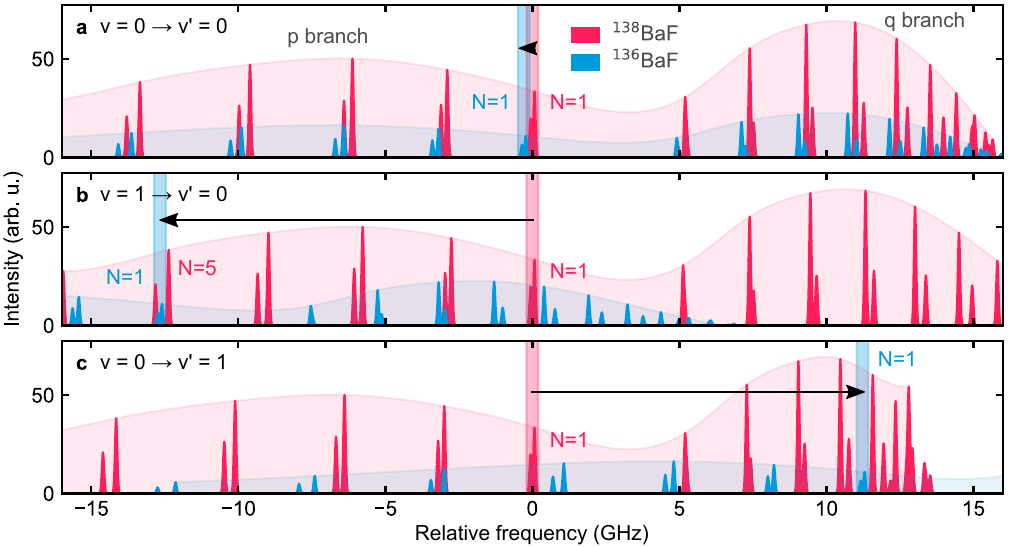}
    \caption{Rovibrational transitions of \baff{136} (blue) and \baff{138} (red) with relevance to optical cycling~\cite{Rockenhaeuser2023}. Other isotopologues contained in the molecular beam have been omitted for clarity. The vertical bars mark the $N=1\rightarrow J'=1/2^+$ cycling transition in BaF (see Fig.~\ref{fig:cyclingscheme}). The horizontal arrows indicate the shift of this transition between the two isotopologues. (a)~Main cooling transition ($\nu=0\rightarrow \nu'=0$), where the shift is reminiscent of the familiar isotope shift in atoms. As its magnitude is comparable to the splitting of the levels involved in the cycling transition, any manipulation of \baff{136} on this transition inherently also leads to off-resonant excitations of \baff{138}. (b)~For the first repumping transition ($\nu=1\rightarrow\nu'=0$), the \baff{136} cooling transition is shifted to much lower frequencies and now overlaps with the $N=5\rightarrow J'=9/2^+$ p-branch transition of \baff{138}. This can lead to spurious short-lived fluorescence from this much more abundant isotopologue, but not to persistent cycling in \baff{138}. When combined with the cooling transition in (a), \baff{138} molecules are thus pumped into the first vibrational state after a few photon scattering events. (c)~Similarly, for the depumping transition ($\nu=0\rightarrow\nu'=1$) used in Raman detection, the cycling transition is shifted to much higher frequencies into the q-branch of \baff{138}, which also does not lead to cycling in \baff{138}. Combining any of these transitions for cooling or detection thus leads to a quasi-closed optical cycle that is highly specific to \baff{136}. Shaded regions are a guide to the eye, indicating the various transition branches. The linestrength of \baff{136} has been increased by a factor of $3$ for better visibility.} 
    \label{fig:isotopeshifts}
\end{figure*}

For vibrations, branching is suppressed through the favorable structure of BaF. Similarly favorable properties are available in a large number of chemically diverse molecular species~\cite{Langen2023,Fitch2021}. In the case of BaF, the ionic bond between the barium and fluorine atoms is formed using one of the two barium valence electrons, while the other optically active electron is primarily located around the barium atom and polarized away from the bond~\cite{Langen2023}. This leads to diagonal Franck-Condon factors, which allow the scattering of thousands of photons, with only one or two vibrational repump lasers that address residual leakage to higher vibrational states. These repumpers drive transitions with $\Delta\nu=\nu'-\nu=-1$ , where $\nu$ and $\nu'$  denote the vibrational quantum numbers in the ground and excited electronic states, respectively (Fig.~\ref{fig:cyclingscheme}a). 

Rotational branching is addressed by exploiting the usual strategy to form a closed optical cycle in alkaline-earth monohalides~\cite{Fitch2021}. Specifically, we drive transitions between the odd parity $N=1$ level in the \gs\, ground state to the even parity $J=1/2^+$ level in the \exs\, state. Here, $N$ denotes the rotational quantum number, $J$ the total angular momentum, and the $+$ in the exponent a state with positive parity. As these states represent the lowest odd (even) parity state in the ground (excited) electronic states, selection rules effectively suppress branching into other ground rotational states. 

Finally, both ground and excited states split further into several hyperfine levels, which need to be addressed using laser sidebands, either resonantly to maximize fluorescence for detection, or in a blue-detuned optical molasses scheme to realize Sisyphus laser cooling forces (Fig.~\ref{fig:cyclingscheme}b). In both processes, dark states are remixed via Larmor precession using an external magnetic field. 

In addition to this regular quasi-closed cycle, which combines the main cooling transition ($\nu=0\rightarrow\nu'=0$) with repumpers ($\Delta\nu=-1$) (Fig.~\ref{fig:cyclingscheme}c), for the detection of the molecules, it is also possible to realize a quasi-closed optical cycle by combining repumpers with depumpers ($\Delta\nu=+1$). The advantage of this Raman scheme is that excitation and fluorescence light are spectrally well separated, facilitating direct imaging of the molecular beam even for rare isotopologues~\cite{Rockenhaeuser2023,Shaw2021}  (Fig.~\ref{fig:cyclingscheme}d).

\section{Realizing isotopologue selectivity}

Since all even, bosonic isotopologues of BaF share a similar level structure, the laser cooling and detection schemes, in principle, are agnostic to the isotopologue used. However, in contrast to atoms, the complexity of the molecular structure renders the navigation of the combined spectra of multiple isotopologues non-trivial. 
While optical cycling in different isotopologues of the same molecular beam has recently been realized in YbOH~\cite{Zeng2023}, to our knowledge, laser cooling of different isotopologues has so far not been achieved.

Compared to atomic laser cooling, where isotope shifts can easily be addressed by adjusting the laser frequencies, molecular laser cooling inherently requires sidebands to address all hyperfine levels at once~\cite{Fitch2020} (see Fig.~\ref{fig:cyclingscheme}b). For BaF, and many other species suitable for laser cooling, isotope shifts of the cooling transitions are comparable to the hyperfine splittings, and thus off-resonant excitations of other isotopologues can often not be avoided. 

For example, when addressing the \baff{136} isotopologue on the usual cycling transition, off-resonant excitations may result in fluorescence being emitted from the more abundant \baff{138} (Fig.~\ref{fig:isotopeshifts}). This fluorescence is indistinguishable from the much weaker \baff{136} fluorescence, making accurate detection of \baff{136} challenging. In turn, when addressing \baff{138} with additional lasers, for example to selectively remove these molecules from the beam, the optical cycle of \baff{136} can easily be disrupted, rendering cooling inefficient. Moreover, in addition to frequency broadened laser spectra, molecular laser cooling also requires higher powers than atomic laser cooling due to the multi-level nature of the cooling transitions used in molecules~\cite{Fitch2021}. The resulting broadening further exacerbates the off-resonant drive of unwanted transitions. 

The challenges of off-resonant excitations can be mitigated by the fact that any sustained optical cycling, either for cooling or detection, involves different vibrational transitions  for cooling, repumping and depumping. This is illustrated in Fig.~\ref{fig:isotopeshifts}, where we outline how the relevant cycling transitions for cooling, repumping and depumping each show an isotope shift that is unique in both direction and magnitude. Any cooling or detection employing a combination of these transitions will therefore make the optical cycle highly specific to a particular isotopologue, while pumping other isotopologues to select dark states. From these dark states, their population can either be recovered, or they can be selectively removed, e.g. by ionization, to obtain beams free of contamination from certain isotopologues. This specificity holds even when spectra of different isotopologues overlap and when laser sidebands off-resonantly excite unwanted nearby transitions. If photon cycling is sustained for long enough, laser cooling and detection will thus only be realized for one specific molecular isotopologue, despite the spectral overlap of the cooling light with other isotopologues.  

\begin{figure}[tb]
    \centering
    \includegraphics[width=0.99\columnwidth]{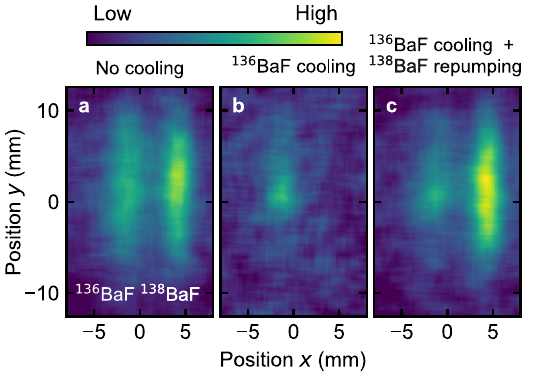}
    \caption{Isotopologue-selective manipulation and cooling. We image both \baff{136} (left) and \baff{138} (right) in parallel using two imaging laser beams. (a) When no cooling light is applied, both transverse molecular profiles show the same width, corresponding to the same transverse temperature of around $100\,$mK. (b) Cooling of \baff{136} results in an accumulation of molecules in the coldest part of the beam, with temperatures well below $2.5\,$mK, while off-resonantly pumping \baff{138} molecules predominantly into the first vibrational state, where they are invisible to the imaging process. (c) Adding a repumping laser for \baff{138} in the imaging region recovers these molecules, demonstrating that they have been unaffected by the cooling of \baff{136} (see also Appendix). Molecular beam direction is from right to left. Images are averaged over $200$ experimental realizations. The colors scale represents molecular signal strength, normalized to the \baff{136} cooling signal. The respective signals reflect both the scattering rates and the natural abundances of the individual isotopologues and can therefore be varied almost at will by using different imaging laser powers.}
    \label{fig:statecontrol}
\end{figure}

We illustrate this principle in Fig.~\ref{fig:statecontrol}, where we image both \baff{136} and \baff{138} in the same molecular beam in parallel. For this, we use two distinct detection laser beams for each isotopologue and image their respective fluorescence together onto our camera. Due to the lower abundance of \baff{136} we employ high-fidelity Raman imaging for this isotopologue, while \baff{138} is imaged using conventional cycling on the cooling transition. 

When no cooling light is applied to the molecular beam, this procedure yields two molecular beam profiles of equal transverse width, and hence, identical temperature (Fig.~\ref{fig:statecontrol}a). In a second step (Fig.~\ref{fig:statecontrol}b), we apply light that is suitable to transversally cool \baff{136} in the interaction region, which results in a strong reduction in transverse velocity of the corresponding molecular beam profile (see also Fig.~\ref{fig:cooling}). As discussed above, the strong cooling light off-resonantly drives transitions in \baff{138}, pumping these molecules predominantly into the first vibrational state after only a few cycles. While the repumping light applied in the cooling region is suitable to address such branching in \baff{136}, it is far detuned from the repumping transitions in \baff{138}. Therefore, no significant forces are realized and, in addition, the corresponding molecules are no longer visible in the image. Residual off-resonant fluorescence of \baff{136} molecules induced by the \baff{138} imaging light can be neglected due to the very low power employed and the low abundance of \baff{136}. Finally, adding a suitable repumper for \baff{138} in the imaging region fully recovers the molecules (Fig.~\ref{fig:statecontrol}c), demonstrating that quantum state control is maintained throughout the experiment. As expected from the arguments above, the recovered \baff{138} molecules show no significant effect of cooling, next to the at least two orders of magnitude colder \baff{136} molecules in the same molecular beam and recorded in the same image. 

\section{Conclusion}
We have demonstrated laser cooling of \baff{136} molecules and their manipulation within a molecular beam containing many naturally occurring BaF isotopologues. Combining the technique with isotopologue-selective ionization, deflection or other types of laser cooling forces is straightforward, and switching between isotopologues can be realized on millisecond timescales by a simple change of laser frequencies. As previously discussed for YbF~\cite{Alauze2021} and \baff{138}~\cite{Rockenhaeuser2024}, the efficiency of the transverse laser cooling can be further enhanced by applying it in both transverse directions. Given the large number of molecular species that are suitable for laser cooling, this could yield intense isotopologue-separated beams of many, chemically diverse species, including also larger polyatomic molecules~\cite{Augenbraun2023}, which can e.g. be useful for cross-beam collision experiments. Furthermore, our technique could provide beams of \baff{135} and \baff{137} for studies of nuclear parity violation~\cite{Altuntas2018,Kogel2021}, or of radioactive \baff{133} and the recently studied isotopologues of the RaF molecule~\cite{GarciaRuiz2020,Udrescu2024}, which feature significant symmetry-violating nuclear moments. Looking further ahead, our proof-of-principle demonstration can be combined with established methods for magneto-optical and dipole trapping~\cite{Fitch2020}, to achieve long-lived samples and manipulation of isotopologues with extremely low abundance. This paves the way for the use of molecules in trace gas analysis, precision measurement applications and isotope-selective chemistry.

\section*{Acknowledgments}
We are indebted to Tilman Pfau for generous support. We thank Bas van der Meeraker for discussions. This project has received funding from the European Research Council (ERC) under the European Union’s Horizon 2020 research and innovation programme (Grant agreements No. 949431), the RiSC programme of the Ministry of Science, Research and Arts Baden-W\"urttemberg and Carl Zeiss Foundation.\\

\section*{Appendix}
\subsection{Experimental implementation}
The implementation of laser cooling requires multiple laser frequencies for cooling and imaging of the molecules~(Fig.~\ref{fig:cyclingscheme}). We derive these frequencies from multiple diode laser systems, which are frequency-stabilized using a transfer cavity~\cite{Pultinevicius2023}. As outlined in the main text, addressing of the hyperfine structure of the molecules requires the creation of additional sidebands for all lasers. For the cooling laser, we use serrodyning using a fiber electro-optical modulator (EOM) to create time-sequenced, optimized sideband spectra~\cite{Rockenhaeuser2024}. For repumping and imaging, we use conventional free-space EOMs to imprint permanent sidebands. All laser systems are subsequently amplified using tapered amplifiers. Typical powers are $65\,$mW (cooler) and $50\,$mW (repumper) for each of the two beams forming the standing wave in the cooling region. As in our previous work, these beams are retroreflected $17$ times~\cite{Rockenhaeuser2024}. For the conventional imaging of \baff{138} we use a single beam containing $150\,$\textmu W of cooling and $120\,$mW of repumping power. Here, the low cooling power is chosen both to minimize off-resonant excitations of \baff{136} molecules during the \baff{138} imaging, and to reduce spurious laser scatter from reflections on the vacuum chamber walls and windows. For Raman imaging of \baff{136}, we use $120\,$mW of depumping and $150\,$mW of repumping power. Laser beams have a diameter of $3.1\,$mm in the transverse cooling region, and $1\,$mm in the imaging region. 

The molecular beam employed has a mean forward velocity of around $170\,$m/s and the transverse velocity spread is around $\pm2.5\,$m/s or, equivalently, around $100\,$mK. Following collimation, the beam can be manipulated using lasers in a $22\,$cm long interaction region.  Subsequently, the molecules expand freely over a distance of $15\,$cm, before they are imaged using laser-induced fluorescence. Further details about the experimental setup have been described in Ref.~\cite{Rockenhaeuser2024}. 

\subsection{Laser cooling mechanism}
The laser cooling mechanism employed in our work is magnetically assisted Sisyphus cooling~\cite{Rockenhaeuser2024}. It relies on the formation of an intensity standing wave in the cooling region, which periodically modulates the energy of bright magnetic sublevels via the AC Stark effect. Molecules travel up the resulting energy hills, loosing kinetic energy in the process. Near the top, the cooling light pumps them into dark magnetic sublevels that do not couple to the cooling laser light. Subsequently, these dark states are remixed with the bright states via Larmor precession induced by applying a suitable magnetic field. We note that the principles making this process isotopologue selective are not restricted to this particular laser cooling mechanism, but can be applied to any other process that relies on the repeated scattering of photons. 

\subsection{Transverse profile of \baff{138}}
\begin{figure}[htb]
    \centering
    \includegraphics[width=\columnwidth]{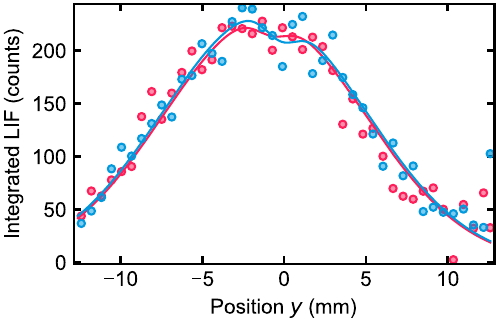}
    \caption{Transverse molecular distribution of \baff{138}, the most abundant isotopologue in the molecular beam. As also shown in Fig.~\ref{fig:statecontrol}, the profiles remain unaffected by the manipulation of \baff{136}. Different colors correspond to heating (red detuned, red data points) and cooling (blue detuned, blue datapoints) for \baff{136}.}
    \label{fig:lineprofiles138}
\end{figure}

In Fig.~\ref{fig:lineprofiles138}, we show the transverse molecular beam profiles of the \baff{138} isotopologue for two parameters of the \baff{136} cooling light. While the latter light results in clear signatures of cooling and heating for the \baff{136} isotopologue (see Fig.~\ref{fig:cooling}), no changes are observed for \baff{138}. This further highlights how the cooling process only acts on one particular isotopologue. \\

\bibliography{biblio}

\begin{thebibliography}{70}%
\makeatletter
\providecommand \@ifxundefined [1]{%
 \@ifx{#1\undefined}
}%
\providecommand \@ifnum [1]{%
 \ifnum #1\expandafter \@firstoftwo
 \else \expandafter \@secondoftwo
 \fi
}%
\providecommand \@ifx [1]{%
 \ifx #1\expandafter \@firstoftwo
 \else \expandafter \@secondoftwo
 \fi
}%
\providecommand \natexlab [1]{#1}%
\providecommand \enquote  [1]{``#1''}%
\providecommand \bibnamefont  [1]{#1}%
\providecommand \bibfnamefont [1]{#1}%
\providecommand \citenamefont [1]{#1}%
\providecommand \href@noop [0]{\@secondoftwo}%
\providecommand \href [0]{\begingroup \@sanitize@url \@href}%
\providecommand \@href[1]{\@@startlink{#1}\@@href}%
\providecommand \@@href[1]{\endgroup#1\@@endlink}%
\providecommand \@sanitize@url [0]{\catcode `\\12\catcode `\$12\catcode
  `\&12\catcode `\#12\catcode `\^12\catcode `\_12\catcode `\%12\relax}%
\providecommand \@@startlink[1]{}%
\providecommand \@@endlink[0]{}%
\providecommand \url  [0]{\begingroup\@sanitize@url \@url }%
\providecommand \@url [1]{\endgroup\@href {#1}{\urlprefix }}%
\providecommand \urlprefix  [0]{URL }%
\providecommand \Eprint [0]{\href }%
\providecommand \doibase [0]{https://doi.org/}%
\providecommand \selectlanguage [0]{\@gobble}%
\providecommand \bibinfo  [0]{\@secondoftwo}%
\providecommand \bibfield  [0]{\@secondoftwo}%
\providecommand \translation [1]{[#1]}%
\providecommand \BibitemOpen [0]{}%
\providecommand \bibitemStop [0]{}%
\providecommand \bibitemNoStop [0]{.\EOS\space}%
\providecommand \EOS [0]{\spacefactor3000\relax}%
\providecommand \BibitemShut  [1]{\csname bibitem#1\endcsname}%
\let\auto@bib@innerbib\@empty
\bibitem [{\citenamefont {Poli}\ \emph {et~al.}(2005)\citenamefont {Poli},
  \citenamefont {Drullinger}, \citenamefont {Ferrari}, \citenamefont
  {L\'eonard}, \citenamefont {Sorrentino},\ and\ \citenamefont
  {Tino}}]{Poli2005}%
  \BibitemOpen
  \bibfield  {author} {\bibinfo {author} {\bibfnamefont {N.}~\bibnamefont
  {Poli}}, \bibinfo {author} {\bibfnamefont {R.~E.}\ \bibnamefont
  {Drullinger}}, \bibinfo {author} {\bibfnamefont {G.}~\bibnamefont {Ferrari}},
  \bibinfo {author} {\bibfnamefont {J.}~\bibnamefont {L\'eonard}}, \bibinfo
  {author} {\bibfnamefont {F.}~\bibnamefont {Sorrentino}},\ and\ \bibinfo
  {author} {\bibfnamefont {G.~M.}\ \bibnamefont {Tino}},\ }\bibfield  {title}
  {\bibinfo {title} {Cooling and trapping of ultracold strontium isotopic
  mixtures},\ }\href {https://doi.org/10.1103/PhysRevA.71.061403} {\bibfield
  {journal} {\bibinfo  {journal} {Phys. Rev. A}\ }\textbf {\bibinfo {volume}
  {71}},\ \bibinfo {pages} {061403} (\bibinfo {year} {2005})}\BibitemShut
  {NoStop}%
\bibitem [{\citenamefont {Papp}\ \emph {et~al.}(2008)\citenamefont {Papp},
  \citenamefont {Pino},\ and\ \citenamefont {Wieman}}]{Papp2008}%
  \BibitemOpen
  \bibfield  {author} {\bibinfo {author} {\bibfnamefont {S.~B.}\ \bibnamefont
  {Papp}}, \bibinfo {author} {\bibfnamefont {J.~M.}\ \bibnamefont {Pino}},\
  and\ \bibinfo {author} {\bibfnamefont {C.~E.}\ \bibnamefont {Wieman}},\
  }\bibfield  {title} {\bibinfo {title} {Tunable miscibility in a dual-species
  {Bose-Einstein} condensate},\ }\href
  {https://doi.org/10.1103/PhysRevLett.101.040402} {\bibfield  {journal}
  {\bibinfo  {journal} {Phys. Rev. Lett.}\ }\textbf {\bibinfo {volume} {101}},\
  \bibinfo {pages} {040402} (\bibinfo {year} {2008})}\BibitemShut {NoStop}%
\bibitem [{\citenamefont {Mickelson}\ \emph {et~al.}(2010)\citenamefont
  {Mickelson}, \citenamefont {Martinez~de Escobar}, \citenamefont {Yan},
  \citenamefont {DeSalvo},\ and\ \citenamefont {Killian}}]{Mickelson2010}%
  \BibitemOpen
  \bibfield  {author} {\bibinfo {author} {\bibfnamefont {P.~G.}\ \bibnamefont
  {Mickelson}}, \bibinfo {author} {\bibfnamefont {Y.~N.}\ \bibnamefont
  {Martinez~de Escobar}}, \bibinfo {author} {\bibfnamefont {M.}~\bibnamefont
  {Yan}}, \bibinfo {author} {\bibfnamefont {B.~J.}\ \bibnamefont {DeSalvo}},\
  and\ \bibinfo {author} {\bibfnamefont {T.~C.}\ \bibnamefont {Killian}},\
  }\bibfield  {title} {\bibinfo {title} {Bose-{Einstein} condensation of
  $^{88}\mathrm{Sr}$ through sympathetic cooling with $^{87}\mathrm{Sr}$},\
  }\href {https://doi.org/10.1103/PhysRevA.81.051601} {\bibfield  {journal}
  {\bibinfo  {journal} {Phys. Rev. A}\ }\textbf {\bibinfo {volume} {81}},\
  \bibinfo {pages} {051601} (\bibinfo {year} {2010})}\BibitemShut {NoStop}%
\bibitem [{\citenamefont {Schreck}\ \emph {et~al.}(2001)\citenamefont
  {Schreck}, \citenamefont {Khaykovich}, \citenamefont {Corwin}, \citenamefont
  {Ferrari}, \citenamefont {Bourdel}, \citenamefont {Cubizolles},\ and\
  \citenamefont {Salomon}}]{Schreck2001}%
  \BibitemOpen
  \bibfield  {author} {\bibinfo {author} {\bibfnamefont {F.}~\bibnamefont
  {Schreck}}, \bibinfo {author} {\bibfnamefont {L.}~\bibnamefont {Khaykovich}},
  \bibinfo {author} {\bibfnamefont {K.~L.}\ \bibnamefont {Corwin}}, \bibinfo
  {author} {\bibfnamefont {G.}~\bibnamefont {Ferrari}}, \bibinfo {author}
  {\bibfnamefont {T.}~\bibnamefont {Bourdel}}, \bibinfo {author} {\bibfnamefont
  {J.}~\bibnamefont {Cubizolles}},\ and\ \bibinfo {author} {\bibfnamefont
  {C.}~\bibnamefont {Salomon}},\ }\bibfield  {title} {\bibinfo {title}
  {Quasipure {Bose-Einstein} condensate immersed in a {Fermi} sea},\ }\href
  {https://doi.org/10.1103/PhysRevLett.87.080403} {\bibfield  {journal}
  {\bibinfo  {journal} {Phys. Rev. Lett.}\ }\textbf {\bibinfo {volume} {87}},\
  \bibinfo {pages} {080403} (\bibinfo {year} {2001})}\BibitemShut {NoStop}%
\bibitem [{\citenamefont {Truscott}\ \emph {et~al.}(2001)\citenamefont
  {Truscott}, \citenamefont {Strecker}, \citenamefont {McAlexander},
  \citenamefont {Partridge},\ and\ \citenamefont {Hulet}}]{Truscott2001}%
  \BibitemOpen
  \bibfield  {author} {\bibinfo {author} {\bibfnamefont {A.~G.}\ \bibnamefont
  {Truscott}}, \bibinfo {author} {\bibfnamefont {K.~E.}\ \bibnamefont
  {Strecker}}, \bibinfo {author} {\bibfnamefont {W.~I.}\ \bibnamefont
  {McAlexander}}, \bibinfo {author} {\bibfnamefont {G.~B.}\ \bibnamefont
  {Partridge}},\ and\ \bibinfo {author} {\bibfnamefont {R.~G.}\ \bibnamefont
  {Hulet}},\ }\bibfield  {title} {\bibinfo {title} {Observation of {Fermi}
  pressure in a gas of trapped atoms},\ }\href
  {https://doi.org/10.1126/science.1059318} {\bibfield  {journal} {\bibinfo
  {journal} {Science}\ }\textbf {\bibinfo {volume} {291}},\ \bibinfo {pages}
  {2570} (\bibinfo {year} {2001})}\BibitemShut {NoStop}%
\bibitem [{\citenamefont {Kato}\ \emph {et~al.}(2017)\citenamefont {Kato},
  \citenamefont {Wang}, \citenamefont {Kobayashi}, \citenamefont {Julienne},\
  and\ \citenamefont {Inouye}}]{Kato2017}%
  \BibitemOpen
  \bibfield  {author} {\bibinfo {author} {\bibfnamefont {K.}~\bibnamefont
  {Kato}}, \bibinfo {author} {\bibfnamefont {Y.}~\bibnamefont {Wang}}, \bibinfo
  {author} {\bibfnamefont {J.}~\bibnamefont {Kobayashi}}, \bibinfo {author}
  {\bibfnamefont {P.~S.}\ \bibnamefont {Julienne}},\ and\ \bibinfo {author}
  {\bibfnamefont {S.}~\bibnamefont {Inouye}},\ }\bibfield  {title} {\bibinfo
  {title} {Isotopic shift of atom-dimer efimov resonances in {K-Rb} mixtures:
  Critical effect of multichannel feshbach physics},\ }\href
  {https://doi.org/10.1103/PhysRevLett.118.163401} {\bibfield  {journal}
  {\bibinfo  {journal} {Phys. Rev. Lett.}\ }\textbf {\bibinfo {volume} {118}},\
  \bibinfo {pages} {163401} (\bibinfo {year} {2017})}\BibitemShut {NoStop}%
\bibitem [{\citenamefont {Tanzi}\ \emph {et~al.}(2018)\citenamefont {Tanzi},
  \citenamefont {Cabrera}, \citenamefont {Sanz}, \citenamefont {Cheiney},
  \citenamefont {Tomza},\ and\ \citenamefont {Tarruell}}]{Tanzi2018}%
  \BibitemOpen
  \bibfield  {author} {\bibinfo {author} {\bibfnamefont {L.}~\bibnamefont
  {Tanzi}}, \bibinfo {author} {\bibfnamefont {C.~R.}\ \bibnamefont {Cabrera}},
  \bibinfo {author} {\bibfnamefont {J.}~\bibnamefont {Sanz}}, \bibinfo {author}
  {\bibfnamefont {P.}~\bibnamefont {Cheiney}}, \bibinfo {author} {\bibfnamefont
  {M.}~\bibnamefont {Tomza}},\ and\ \bibinfo {author} {\bibfnamefont
  {L.}~\bibnamefont {Tarruell}},\ }\bibfield  {title} {\bibinfo {title}
  {Feshbach resonances in potassium {Bose-Bose} mixtures},\ }\href
  {https://doi.org/10.1103/PhysRevA.98.062712} {\bibfield  {journal} {\bibinfo
  {journal} {Phys. Rev. A}\ }\textbf {\bibinfo {volume} {98}},\ \bibinfo
  {pages} {062712} (\bibinfo {year} {2018})}\BibitemShut {NoStop}%
\bibitem [{\citenamefont {Durastante}\ \emph {et~al.}(2020)\citenamefont
  {Durastante}, \citenamefont {Politi}, \citenamefont {Sohmen}, \citenamefont
  {Ilzh\"ofer}, \citenamefont {Mark}, \citenamefont {Norcia},\ and\
  \citenamefont {Ferlaino}}]{Durastante2020}%
  \BibitemOpen
  \bibfield  {author} {\bibinfo {author} {\bibfnamefont {G.}~\bibnamefont
  {Durastante}}, \bibinfo {author} {\bibfnamefont {C.}~\bibnamefont {Politi}},
  \bibinfo {author} {\bibfnamefont {M.}~\bibnamefont {Sohmen}}, \bibinfo
  {author} {\bibfnamefont {P.}~\bibnamefont {Ilzh\"ofer}}, \bibinfo {author}
  {\bibfnamefont {M.~J.}\ \bibnamefont {Mark}}, \bibinfo {author}
  {\bibfnamefont {M.~A.}\ \bibnamefont {Norcia}},\ and\ \bibinfo {author}
  {\bibfnamefont {F.}~\bibnamefont {Ferlaino}},\ }\bibfield  {title} {\bibinfo
  {title} {Feshbach resonances in an erbium-dysprosium dipolar mixture},\
  }\href {https://doi.org/10.1103/PhysRevA.102.033330} {\bibfield  {journal}
  {\bibinfo  {journal} {Phys. Rev. A}\ }\textbf {\bibinfo {volume} {102}},\
  \bibinfo {pages} {033330} (\bibinfo {year} {2020})}\BibitemShut {NoStop}%
\bibitem [{\citenamefont {Bernhardt}\ \emph {et~al.}(1974)\citenamefont
  {Bernhardt}, \citenamefont {Duerre}, \citenamefont {Simpson},\ and\
  \citenamefont {Wood}}]{Bernhardt1974}%
  \BibitemOpen
  \bibfield  {author} {\bibinfo {author} {\bibfnamefont {A.~F.}\ \bibnamefont
  {Bernhardt}}, \bibinfo {author} {\bibfnamefont {D.~E.}\ \bibnamefont
  {Duerre}}, \bibinfo {author} {\bibfnamefont {J.~R.}\ \bibnamefont
  {Simpson}},\ and\ \bibinfo {author} {\bibfnamefont {L.~L.}\ \bibnamefont
  {Wood}},\ }\bibfield  {title} {\bibinfo {title} {{Separation of isotopes by
  laser deflection of atomic beam. I. Barium}},\ }\href
  {https://doi.org/10.1063/1.1655333} {\bibfield  {journal} {\bibinfo
  {journal} {Applied Physics Letters}\ }\textbf {\bibinfo {volume} {25}},\
  \bibinfo {pages} {617} (\bibinfo {year} {1974})}\BibitemShut {NoStop}%
\bibitem [{\citenamefont {Lu}\ \emph {et~al.}(1994)\citenamefont {Lu},
  \citenamefont {Bowers}, \citenamefont {Freedman}, \citenamefont {Fujikawa},
  \citenamefont {Mortara}, \citenamefont {Shang}, \citenamefont {Coulter},\
  and\ \citenamefont {Young}}]{Lu1994}%
  \BibitemOpen
  \bibfield  {author} {\bibinfo {author} {\bibfnamefont {Z.-T.}\ \bibnamefont
  {Lu}}, \bibinfo {author} {\bibfnamefont {C.}~\bibnamefont {Bowers}}, \bibinfo
  {author} {\bibfnamefont {S.~J.}\ \bibnamefont {Freedman}}, \bibinfo {author}
  {\bibfnamefont {B.~K.}\ \bibnamefont {Fujikawa}}, \bibinfo {author}
  {\bibfnamefont {J.~L.}\ \bibnamefont {Mortara}}, \bibinfo {author}
  {\bibfnamefont {S.-Q.}\ \bibnamefont {Shang}}, \bibinfo {author}
  {\bibfnamefont {K.~P.}\ \bibnamefont {Coulter}},\ and\ \bibinfo {author}
  {\bibfnamefont {L.}~\bibnamefont {Young}},\ }\bibfield  {title} {\bibinfo
  {title} {Laser trapping of short-lived radioactive isotopes},\ }\href
  {https://doi.org/10.1103/PhysRevLett.72.3791} {\bibfield  {journal} {\bibinfo
   {journal} {Phys. Rev. Lett.}\ }\textbf {\bibinfo {volume} {72}},\ \bibinfo
  {pages} {3791} (\bibinfo {year} {1994})}\BibitemShut {NoStop}%
\bibitem [{\citenamefont {Gwinner}\ \emph {et~al.}(1994)\citenamefont
  {Gwinner}, \citenamefont {Behr}, \citenamefont {Cahn}, \citenamefont {Ghosh},
  \citenamefont {Orozco}, \citenamefont {Sprouse},\ and\ \citenamefont
  {Xu}}]{Gwinner1994}%
  \BibitemOpen
  \bibfield  {author} {\bibinfo {author} {\bibfnamefont {G.}~\bibnamefont
  {Gwinner}}, \bibinfo {author} {\bibfnamefont {J.~A.}\ \bibnamefont {Behr}},
  \bibinfo {author} {\bibfnamefont {S.~B.}\ \bibnamefont {Cahn}}, \bibinfo
  {author} {\bibfnamefont {A.}~\bibnamefont {Ghosh}}, \bibinfo {author}
  {\bibfnamefont {L.~A.}\ \bibnamefont {Orozco}}, \bibinfo {author}
  {\bibfnamefont {G.~D.}\ \bibnamefont {Sprouse}},\ and\ \bibinfo {author}
  {\bibfnamefont {F.}~\bibnamefont {Xu}},\ }\bibfield  {title} {\bibinfo
  {title} {Magneto-optic trapping of radioactive $^{79}\mathrm{Rb}$},\ }\href
  {https://doi.org/10.1103/PhysRevLett.72.3795} {\bibfield  {journal} {\bibinfo
   {journal} {Phys. Rev. Lett.}\ }\textbf {\bibinfo {volume} {72}},\ \bibinfo
  {pages} {3795} (\bibinfo {year} {1994})}\BibitemShut {NoStop}%
\bibitem [{\citenamefont {Sisarian}\ \emph {et~al.}(1996)\citenamefont
  {Sisarian}, \citenamefont {Ghosh}, \citenamefont {Gwinner}, \citenamefont
  {Orozco}, \citenamefont {Sprouse},\ and\ \citenamefont
  {Voytas}}]{Simsarian1996}%
  \BibitemOpen
  \bibfield  {author} {\bibinfo {author} {\bibfnamefont {J.~E.}\ \bibnamefont
  {Sisarian}}, \bibinfo {author} {\bibfnamefont {A.}~\bibnamefont {Ghosh}},
  \bibinfo {author} {\bibfnamefont {G.}~\bibnamefont {Gwinner}}, \bibinfo
  {author} {\bibfnamefont {L.~A.}\ \bibnamefont {Orozco}}, \bibinfo {author}
  {\bibfnamefont {G.~D.}\ \bibnamefont {Sprouse}},\ and\ \bibinfo {author}
  {\bibfnamefont {P.~A.}\ \bibnamefont {Voytas}},\ }\bibfield  {title}
  {\bibinfo {title} {Magneto-optic trapping of ${}^{210}${Fr}},\ }\href
  {https://doi.org/10.1103/PhysRevLett.76.3522} {\bibfield  {journal} {\bibinfo
   {journal} {Phys. Rev. Lett.}\ }\textbf {\bibinfo {volume} {76}},\ \bibinfo
  {pages} {3522} (\bibinfo {year} {1996})}\BibitemShut {NoStop}%
\bibitem [{\citenamefont {Lu}\ \emph {et~al.}(1997)\citenamefont {Lu},
  \citenamefont {Corwin}, \citenamefont {Vogel}, \citenamefont {Wieman},
  \citenamefont {Dinneen}, \citenamefont {Maddi},\ and\ \citenamefont
  {Gould}}]{Lu1997}%
  \BibitemOpen
  \bibfield  {author} {\bibinfo {author} {\bibfnamefont {Z.-T.}\ \bibnamefont
  {Lu}}, \bibinfo {author} {\bibfnamefont {K.~L.}\ \bibnamefont {Corwin}},
  \bibinfo {author} {\bibfnamefont {K.~R.}\ \bibnamefont {Vogel}}, \bibinfo
  {author} {\bibfnamefont {C.~E.}\ \bibnamefont {Wieman}}, \bibinfo {author}
  {\bibfnamefont {T.~P.}\ \bibnamefont {Dinneen}}, \bibinfo {author}
  {\bibfnamefont {J.}~\bibnamefont {Maddi}},\ and\ \bibinfo {author}
  {\bibfnamefont {H.}~\bibnamefont {Gould}},\ }\bibfield  {title} {\bibinfo
  {title} {Efficient collection of ${}^{221}${Fr} into a vapor cell
  magneto-optical trap},\ }\href {https://doi.org/10.1103/PhysRevLett.79.994}
  {\bibfield  {journal} {\bibinfo  {journal} {Phys. Rev. Lett.}\ }\textbf
  {\bibinfo {volume} {79}},\ \bibinfo {pages} {994} (\bibinfo {year}
  {1997})}\BibitemShut {NoStop}%
\bibitem [{\citenamefont {Guest}\ \emph {et~al.}(2007)\citenamefont {Guest},
  \citenamefont {Scielzo}, \citenamefont {Ahmad}, \citenamefont {Bailey},
  \citenamefont {Greene}, \citenamefont {Holt}, \citenamefont {Lu},
  \citenamefont {O'Connor},\ and\ \citenamefont {Potterveld}}]{Guest2007}%
  \BibitemOpen
  \bibfield  {author} {\bibinfo {author} {\bibfnamefont {J.~R.}\ \bibnamefont
  {Guest}}, \bibinfo {author} {\bibfnamefont {N.~D.}\ \bibnamefont {Scielzo}},
  \bibinfo {author} {\bibfnamefont {I.}~\bibnamefont {Ahmad}}, \bibinfo
  {author} {\bibfnamefont {K.}~\bibnamefont {Bailey}}, \bibinfo {author}
  {\bibfnamefont {J.~P.}\ \bibnamefont {Greene}}, \bibinfo {author}
  {\bibfnamefont {R.~J.}\ \bibnamefont {Holt}}, \bibinfo {author}
  {\bibfnamefont {Z.-T.}\ \bibnamefont {Lu}}, \bibinfo {author} {\bibfnamefont
  {T.~P.}\ \bibnamefont {O'Connor}},\ and\ \bibinfo {author} {\bibfnamefont
  {D.~H.}\ \bibnamefont {Potterveld}},\ }\bibfield  {title} {\bibinfo {title}
  {Laser trapping of $^{225}\mathrm{Ra}$ and $^{226}\mathrm{Ra}$ with repumping
  by room-temperature blackbody radiation},\ }\href
  {https://doi.org/10.1103/PhysRevLett.98.093001} {\bibfield  {journal}
  {\bibinfo  {journal} {Phys. Rev. Lett.}\ }\textbf {\bibinfo {volume} {98}},\
  \bibinfo {pages} {093001} (\bibinfo {year} {2007})}\BibitemShut {NoStop}%
\bibitem [{\citenamefont {Chen}\ \emph {et~al.}(1999)\citenamefont {Chen},
  \citenamefont {Li}, \citenamefont {Bailey}, \citenamefont {O'Connor},
  \citenamefont {Young},\ and\ \citenamefont {Lu}}]{Chen1999}%
  \BibitemOpen
  \bibfield  {author} {\bibinfo {author} {\bibfnamefont {C.~Y.}\ \bibnamefont
  {Chen}}, \bibinfo {author} {\bibfnamefont {Y.~M.}\ \bibnamefont {Li}},
  \bibinfo {author} {\bibfnamefont {K.}~\bibnamefont {Bailey}}, \bibinfo
  {author} {\bibfnamefont {T.~P.}\ \bibnamefont {O'Connor}}, \bibinfo {author}
  {\bibfnamefont {L.}~\bibnamefont {Young}},\ and\ \bibinfo {author}
  {\bibfnamefont {Z.~T.}\ \bibnamefont {Lu}},\ }\bibfield  {title} {\bibinfo
  {title} {Ultrasensitive isotope trace analyses with a magneto-optical trap},\
  }\href {https://doi.org/10.1126/science.286.5442.1139} {\bibfield  {journal}
  {\bibinfo  {journal} {Science}\ }\textbf {\bibinfo {volume} {286}},\ \bibinfo
  {pages} {1139} (\bibinfo {year} {1999})}\BibitemShut {NoStop}%
\bibitem [{\citenamefont {Hoekstra}\ \emph {et~al.}(2005)\citenamefont
  {Hoekstra}, \citenamefont {Mollema}, \citenamefont {Morgenstern},
  \citenamefont {Wilschut},\ and\ \citenamefont {Hoekstra}}]{Hoekstra2005}%
  \BibitemOpen
  \bibfield  {author} {\bibinfo {author} {\bibfnamefont {S.}~\bibnamefont
  {Hoekstra}}, \bibinfo {author} {\bibfnamefont {A.~K.}\ \bibnamefont
  {Mollema}}, \bibinfo {author} {\bibfnamefont {R.}~\bibnamefont
  {Morgenstern}}, \bibinfo {author} {\bibfnamefont {H.~W.}\ \bibnamefont
  {Wilschut}},\ and\ \bibinfo {author} {\bibfnamefont {R.}~\bibnamefont
  {Hoekstra}},\ }\bibfield  {title} {\bibinfo {title} {Single-atom detection of
  calcium isotopes by atom-trap trace analysis},\ }\href
  {https://doi.org/10.1103/PhysRevA.71.023409} {\bibfield  {journal} {\bibinfo
  {journal} {Phys. Rev. A}\ }\textbf {\bibinfo {volume} {71}},\ \bibinfo
  {pages} {023409} (\bibinfo {year} {2005})}\BibitemShut {NoStop}%
\bibitem [{\citenamefont {Welte}\ \emph {et~al.}(2010)\citenamefont {Welte},
  \citenamefont {Ritterbusch}, \citenamefont {Steinke}, \citenamefont
  {Henrich}, \citenamefont {Aeschbach-Hertig},\ and\ \citenamefont
  {Oberthaler}}]{Welte2010}%
  \BibitemOpen
  \bibfield  {author} {\bibinfo {author} {\bibfnamefont {J.}~\bibnamefont
  {Welte}}, \bibinfo {author} {\bibfnamefont {F.}~\bibnamefont {Ritterbusch}},
  \bibinfo {author} {\bibfnamefont {I.}~\bibnamefont {Steinke}}, \bibinfo
  {author} {\bibfnamefont {M.}~\bibnamefont {Henrich}}, \bibinfo {author}
  {\bibfnamefont {W.}~\bibnamefont {Aeschbach-Hertig}},\ and\ \bibinfo {author}
  {\bibfnamefont {M.~K.}\ \bibnamefont {Oberthaler}},\ }\bibfield  {title}
  {\bibinfo {title} {Towards the realization of atom trap trace analysis for
  39{Ar}},\ }\href {https://doi.org/10.1088/1367-2630/12/6/065031} {\bibfield
  {journal} {\bibinfo  {journal} {New Journal of Physics}\ }\textbf {\bibinfo
  {volume} {12}},\ \bibinfo {pages} {065031} (\bibinfo {year}
  {2010})}\BibitemShut {NoStop}%
\bibitem [{\citenamefont {Scielzo}\ \emph {et~al.}(2004)\citenamefont
  {Scielzo}, \citenamefont {Freedman}, \citenamefont {Fujikawa},\ and\
  \citenamefont {Vetter}}]{Scielzo2004}%
  \BibitemOpen
  \bibfield  {author} {\bibinfo {author} {\bibfnamefont {N.~D.}\ \bibnamefont
  {Scielzo}}, \bibinfo {author} {\bibfnamefont {S.~J.}\ \bibnamefont
  {Freedman}}, \bibinfo {author} {\bibfnamefont {B.~K.}\ \bibnamefont
  {Fujikawa}},\ and\ \bibinfo {author} {\bibfnamefont {P.~A.}\ \bibnamefont
  {Vetter}},\ }\bibfield  {title} {\bibinfo {title} {Measurement of the
  $\ensuremath{\beta}\ensuremath{-}\ensuremath{\nu}$ correlation using
  magneto-optically trapped $^{21}\mathrm{N}\mathrm{a}$},\ }\href
  {https://doi.org/10.1103/PhysRevLett.93.102501} {\bibfield  {journal}
  {\bibinfo  {journal} {Phys. Rev. Lett.}\ }\textbf {\bibinfo {volume} {93}},\
  \bibinfo {pages} {102501} (\bibinfo {year} {2004})}\BibitemShut {NoStop}%
\bibitem [{\citenamefont {Gorelov}\ \emph {et~al.}(2005)\citenamefont
  {Gorelov}, \citenamefont {Melconian}, \citenamefont {Alford}, \citenamefont
  {Ashery}, \citenamefont {Ball}, \citenamefont {Behr}, \citenamefont
  {Bricault}, \citenamefont {D'Auria}, \citenamefont {Deutsch}, \citenamefont
  {Dilling}, \citenamefont {Dombsky}, \citenamefont {Dub\'e}, \citenamefont
  {Fingler}, \citenamefont {Giesen}, \citenamefont {Gl\"uck}, \citenamefont
  {Gu}, \citenamefont {H\"ausser}, \citenamefont {Jackson}, \citenamefont
  {Jennings}, \citenamefont {Pearson}, \citenamefont {Stocki}, \citenamefont
  {Swanson},\ and\ \citenamefont {Trinczek}}]{Gorelov2005}%
  \BibitemOpen
  \bibfield  {author} {\bibinfo {author} {\bibfnamefont {A.}~\bibnamefont
  {Gorelov}}, \bibinfo {author} {\bibfnamefont {D.}~\bibnamefont {Melconian}},
  \bibinfo {author} {\bibfnamefont {W.~P.}\ \bibnamefont {Alford}}, \bibinfo
  {author} {\bibfnamefont {D.}~\bibnamefont {Ashery}}, \bibinfo {author}
  {\bibfnamefont {G.}~\bibnamefont {Ball}}, \bibinfo {author} {\bibfnamefont
  {J.~A.}\ \bibnamefont {Behr}}, \bibinfo {author} {\bibfnamefont {P.~G.}\
  \bibnamefont {Bricault}}, \bibinfo {author} {\bibfnamefont {J.~M.}\
  \bibnamefont {D'Auria}}, \bibinfo {author} {\bibfnamefont {J.}~\bibnamefont
  {Deutsch}}, \bibinfo {author} {\bibfnamefont {J.}~\bibnamefont {Dilling}},
  \bibinfo {author} {\bibfnamefont {M.}~\bibnamefont {Dombsky}}, \bibinfo
  {author} {\bibfnamefont {P.}~\bibnamefont {Dub\'e}}, \bibinfo {author}
  {\bibfnamefont {J.}~\bibnamefont {Fingler}}, \bibinfo {author} {\bibfnamefont
  {U.}~\bibnamefont {Giesen}}, \bibinfo {author} {\bibfnamefont
  {F.}~\bibnamefont {Gl\"uck}}, \bibinfo {author} {\bibfnamefont
  {S.}~\bibnamefont {Gu}}, \bibinfo {author} {\bibfnamefont {O.}~\bibnamefont
  {H\"ausser}}, \bibinfo {author} {\bibfnamefont {K.~P.}\ \bibnamefont
  {Jackson}}, \bibinfo {author} {\bibfnamefont {B.~K.}\ \bibnamefont
  {Jennings}}, \bibinfo {author} {\bibfnamefont {M.~R.}\ \bibnamefont
  {Pearson}}, \bibinfo {author} {\bibfnamefont {T.~J.}\ \bibnamefont {Stocki}},
  \bibinfo {author} {\bibfnamefont {T.~B.}\ \bibnamefont {Swanson}},\ and\
  \bibinfo {author} {\bibfnamefont {M.}~\bibnamefont {Trinczek}},\ }\bibfield
  {title} {\bibinfo {title} {Scalar interaction limits from the
  $\ensuremath{\beta}\mathrm{\text{\ensuremath{-}}}\ensuremath{\nu}$
  correlation of trapped radioactive atoms},\ }\href
  {https://doi.org/10.1103/PhysRevLett.94.142501} {\bibfield  {journal}
  {\bibinfo  {journal} {Phys. Rev. Lett.}\ }\textbf {\bibinfo {volume} {94}},\
  \bibinfo {pages} {142501} (\bibinfo {year} {2005})}\BibitemShut {NoStop}%
\bibitem [{\citenamefont {Mueller}\ \emph {et~al.}(2007)\citenamefont
  {Mueller}, \citenamefont {Sulai}, \citenamefont {Villari}, \citenamefont
  {Alc\'antara-N\'u\~nez}, \citenamefont {Alves-Cond\'e}, \citenamefont
  {Bailey}, \citenamefont {Drake}, \citenamefont {Dubois}, \citenamefont
  {El\'eon}, \citenamefont {Gaubert}, \citenamefont {Holt}, \citenamefont
  {Janssens}, \citenamefont {Lecesne}, \citenamefont {Lu}, \citenamefont
  {O'Connor}, \citenamefont {Saint-Laurent}, \citenamefont {Thomas},\ and\
  \citenamefont {Wang}}]{Mueller2007}%
  \BibitemOpen
  \bibfield  {author} {\bibinfo {author} {\bibfnamefont {P.}~\bibnamefont
  {Mueller}}, \bibinfo {author} {\bibfnamefont {I.~A.}\ \bibnamefont {Sulai}},
  \bibinfo {author} {\bibfnamefont {A.~C.~C.}\ \bibnamefont {Villari}},
  \bibinfo {author} {\bibfnamefont {J.~A.}\ \bibnamefont
  {Alc\'antara-N\'u\~nez}}, \bibinfo {author} {\bibfnamefont {R.}~\bibnamefont
  {Alves-Cond\'e}}, \bibinfo {author} {\bibfnamefont {K.}~\bibnamefont
  {Bailey}}, \bibinfo {author} {\bibfnamefont {G.~W.~F.}\ \bibnamefont
  {Drake}}, \bibinfo {author} {\bibfnamefont {M.}~\bibnamefont {Dubois}},
  \bibinfo {author} {\bibfnamefont {C.}~\bibnamefont {El\'eon}}, \bibinfo
  {author} {\bibfnamefont {G.}~\bibnamefont {Gaubert}}, \bibinfo {author}
  {\bibfnamefont {R.~J.}\ \bibnamefont {Holt}}, \bibinfo {author}
  {\bibfnamefont {R.~V.~F.}\ \bibnamefont {Janssens}}, \bibinfo {author}
  {\bibfnamefont {N.}~\bibnamefont {Lecesne}}, \bibinfo {author} {\bibfnamefont
  {Z.-T.}\ \bibnamefont {Lu}}, \bibinfo {author} {\bibfnamefont {T.~P.}\
  \bibnamefont {O'Connor}}, \bibinfo {author} {\bibfnamefont {M.-G.}\
  \bibnamefont {Saint-Laurent}}, \bibinfo {author} {\bibfnamefont {J.-C.}\
  \bibnamefont {Thomas}},\ and\ \bibinfo {author} {\bibfnamefont {L.-B.}\
  \bibnamefont {Wang}},\ }\bibfield  {title} {\bibinfo {title} {Nuclear charge
  radius of $^{8}\mathrm{He}$},\ }\href
  {https://doi.org/10.1103/PhysRevLett.99.252501} {\bibfield  {journal}
  {\bibinfo  {journal} {Phys. Rev. Lett.}\ }\textbf {\bibinfo {volume} {99}},\
  \bibinfo {pages} {252501} (\bibinfo {year} {2007})}\BibitemShut {NoStop}%
\bibitem [{\citenamefont {Wilschut}\ \emph {et~al.}(2010)\citenamefont
  {Wilschut}, \citenamefont {{van der Hoek}}, \citenamefont {Jungmann},
  \citenamefont {Kruithof}, \citenamefont {Onderwater}, \citenamefont {Santra},
  \citenamefont {Shidling},\ and\ \citenamefont {Willmann}}]{Wilschut2010}%
  \BibitemOpen
  \bibfield  {author} {\bibinfo {author} {\bibfnamefont {H.}~\bibnamefont
  {Wilschut}}, \bibinfo {author} {\bibfnamefont {D.}~\bibnamefont {{van der
  Hoek}}}, \bibinfo {author} {\bibfnamefont {K.}~\bibnamefont {Jungmann}},
  \bibinfo {author} {\bibfnamefont {W.}~\bibnamefont {Kruithof}}, \bibinfo
  {author} {\bibfnamefont {C.}~\bibnamefont {Onderwater}}, \bibinfo {author}
  {\bibfnamefont {B.}~\bibnamefont {Santra}}, \bibinfo {author} {\bibfnamefont
  {P.}~\bibnamefont {Shidling}},\ and\ \bibinfo {author} {\bibfnamefont
  {L.}~\bibnamefont {Willmann}},\ }\bibfield  {title} {\bibinfo {title}
  {$\beta$-decay and the electric dipole moment: searches for time-reversal
  violation in radioactive nuclei and atoms},\ }\href
  {https://doi.org/https://doi.org/10.1016/j.nuclphysa.2010.05.025} {\bibfield
  {journal} {\bibinfo  {journal} {Nuclear Physics A}\ }\textbf {\bibinfo
  {volume} {844}},\ \bibinfo {pages} {143c} (\bibinfo {year} {2010})},\
  \bibinfo {note} {proceedings of the 4th International Symposium on Symmetries
  in Subatomic Physics}\BibitemShut {NoStop}%
\bibitem [{\citenamefont {Martoff}\ \emph {et~al.}(2021)\citenamefont
  {Martoff}, \citenamefont {Granato}, \citenamefont {Palmaccio}, \citenamefont
  {Yu}, \citenamefont {Smith}, \citenamefont {Hudson}, \citenamefont
  {Hamilton}, \citenamefont {Schneider}, \citenamefont {Chang}, \citenamefont
  {Renshaw}, \citenamefont {Malatino}, \citenamefont {Meyers},\ and\
  \citenamefont {Lamichhane}}]{Martoff2021}%
  \BibitemOpen
  \bibfield  {author} {\bibinfo {author} {\bibfnamefont {C.~J.}\ \bibnamefont
  {Martoff}}, \bibinfo {author} {\bibfnamefont {F.}~\bibnamefont {Granato}},
  \bibinfo {author} {\bibfnamefont {V.}~\bibnamefont {Palmaccio}}, \bibinfo
  {author} {\bibfnamefont {X.}~\bibnamefont {Yu}}, \bibinfo {author}
  {\bibfnamefont {P.~F.}\ \bibnamefont {Smith}}, \bibinfo {author}
  {\bibfnamefont {E.~R.}\ \bibnamefont {Hudson}}, \bibinfo {author}
  {\bibfnamefont {P.}~\bibnamefont {Hamilton}}, \bibinfo {author}
  {\bibfnamefont {C.}~\bibnamefont {Schneider}}, \bibinfo {author}
  {\bibfnamefont {E.}~\bibnamefont {Chang}}, \bibinfo {author} {\bibfnamefont
  {A.}~\bibnamefont {Renshaw}}, \bibinfo {author} {\bibfnamefont
  {F.}~\bibnamefont {Malatino}}, \bibinfo {author} {\bibfnamefont {P.~D.}\
  \bibnamefont {Meyers}},\ and\ \bibinfo {author} {\bibfnamefont
  {B.}~\bibnamefont {Lamichhane}},\ }\bibfield  {title} {\bibinfo {title}
  {Hunter: precision massive-neutrino search based on a laser cooled atomic
  source},\ }\href {https://doi.org/10.1088/2058-9565/abdb9b} {\bibfield
  {journal} {\bibinfo  {journal} {Quantum Science and Technology}\ }\textbf
  {\bibinfo {volume} {6}},\ \bibinfo {pages} {024008} (\bibinfo {year}
  {2021})}\BibitemShut {NoStop}%
\bibitem [{\citenamefont {Shuman}\ \emph {et~al.}(2010)\citenamefont {Shuman},
  \citenamefont {Barry},\ and\ \citenamefont {Demille}}]{Shuman2010}%
  \BibitemOpen
  \bibfield  {author} {\bibinfo {author} {\bibfnamefont {E.~S.}\ \bibnamefont
  {Shuman}}, \bibinfo {author} {\bibfnamefont {J.~F.}\ \bibnamefont {Barry}},\
  and\ \bibinfo {author} {\bibfnamefont {D.}~\bibnamefont {Demille}},\
  }\bibfield  {title} {\bibinfo {title} {{Laser cooling of a diatomic
  molecule}},\ }\href {https://doi.org/10.1038/nature09443} {\bibfield
  {journal} {\bibinfo  {journal} {Nature}\ }\textbf {\bibinfo {volume} {467}},\
  \bibinfo {pages} {820} (\bibinfo {year} {2010})},\ \Eprint
  {https://arxiv.org/abs/1103.6004} {arXiv:1103.6004} \BibitemShut {NoStop}%
\bibitem [{\citenamefont {Fitch}\ and\ \citenamefont
  {Tarbutt}(2021)}]{Fitch2021}%
  \BibitemOpen
  \bibfield  {author} {\bibinfo {author} {\bibfnamefont {N.}~\bibnamefont
  {Fitch}}\ and\ \bibinfo {author} {\bibfnamefont {M.}~\bibnamefont
  {Tarbutt}},\ }\bibfield  {title} {\bibinfo {title} {Laser-cooled molecules},\
  }\href {https://doi.org/https://doi.org/10.1016/bs.aamop.2021.04.003}
  {\bibfield  {journal} {\bibinfo  {journal} {Advances In Atomic, Molecular,
  and Optical Physics}\ }\textbf {\bibinfo {volume} {70}},\ \bibinfo {pages}
  {157} (\bibinfo {year} {2021})}\BibitemShut {NoStop}%
\bibitem [{\citenamefont {McCarron}(2018)}]{McCarron2018a}%
  \BibitemOpen
  \bibfield  {author} {\bibinfo {author} {\bibfnamefont {D.}~\bibnamefont
  {McCarron}},\ }\bibfield  {title} {\bibinfo {title} {{Laser cooling and
  trapping molecules}},\ }\href {https://doi.org/10.1088/1361-6455/aadfba}
  {\bibfield  {journal} {\bibinfo  {journal} {Journal of Physics B: Atomic,
  Molecular and Optical Physics}\ }\textbf {\bibinfo {volume} {51}},\ \bibinfo
  {pages} {212001} (\bibinfo {year} {2018})}\BibitemShut {NoStop}%
\bibitem [{\citenamefont {Langen}\ \emph {et~al.}(2024)\citenamefont {Langen},
  \citenamefont {Valtolina}, \citenamefont {Wang},\ and\ \citenamefont
  {Ye}}]{Langen2023}%
  \BibitemOpen
  \bibfield  {author} {\bibinfo {author} {\bibfnamefont {T.}~\bibnamefont
  {Langen}}, \bibinfo {author} {\bibfnamefont {G.}~\bibnamefont {Valtolina}},
  \bibinfo {author} {\bibfnamefont {D.}~\bibnamefont {Wang}},\ and\ \bibinfo
  {author} {\bibfnamefont {J.}~\bibnamefont {Ye}},\ }\bibfield  {title}
  {\bibinfo {title} {{Quantum state manipulation and cooling of ultracold
  molecules}},\ }\href {https://doi.org/10.1038/s41567-024-02423-1} {\bibfield
  {journal} {\bibinfo  {journal} {Nature Physics}\ }\textbf {\bibinfo {volume}
  {20}},\ \bibinfo {pages} {702} (\bibinfo {year} {2024})}\BibitemShut
  {NoStop}%
\bibitem [{\citenamefont {Barry}\ \emph {et~al.}(2014)\citenamefont {Barry},
  \citenamefont {McCarron}, \citenamefont {Norrgard}, \citenamefont
  {Steinecker},\ and\ \citenamefont {Demille}}]{Barry2014}%
  \BibitemOpen
  \bibfield  {author} {\bibinfo {author} {\bibfnamefont {J.~F.}\ \bibnamefont
  {Barry}}, \bibinfo {author} {\bibfnamefont {D.~J.}\ \bibnamefont {McCarron}},
  \bibinfo {author} {\bibfnamefont {E.~B.}\ \bibnamefont {Norrgard}}, \bibinfo
  {author} {\bibfnamefont {M.~H.}\ \bibnamefont {Steinecker}},\ and\ \bibinfo
  {author} {\bibfnamefont {D.}~\bibnamefont {Demille}},\ }\bibfield  {title}
  {\bibinfo {title} {{Magneto-optical trapping of a diatomic molecule}},\
  }\href {https://doi.org/10.1038/nature13634} {\bibfield  {journal} {\bibinfo
  {journal} {Nature}\ }\textbf {\bibinfo {volume} {512}},\ \bibinfo {pages}
  {286} (\bibinfo {year} {2014})},\ \Eprint {https://arxiv.org/abs/1404.5680}
  {arXiv:1404.5680} \BibitemShut {NoStop}%
\bibitem [{\citenamefont {Collopy}\ \emph {et~al.}(2018)\citenamefont
  {Collopy}, \citenamefont {Ding}, \citenamefont {Wu}, \citenamefont
  {Finneran}, \citenamefont {Anderegg}, \citenamefont {Augenbraun},
  \citenamefont {Doyle},\ and\ \citenamefont {Ye}}]{Collopy2018}%
  \BibitemOpen
  \bibfield  {author} {\bibinfo {author} {\bibfnamefont {A.~L.}\ \bibnamefont
  {Collopy}}, \bibinfo {author} {\bibfnamefont {S.}~\bibnamefont {Ding}},
  \bibinfo {author} {\bibfnamefont {Y.}~\bibnamefont {Wu}}, \bibinfo {author}
  {\bibfnamefont {I.~A.}\ \bibnamefont {Finneran}}, \bibinfo {author}
  {\bibfnamefont {L.}~\bibnamefont {Anderegg}}, \bibinfo {author}
  {\bibfnamefont {B.~L.}\ \bibnamefont {Augenbraun}}, \bibinfo {author}
  {\bibfnamefont {J.~M.}\ \bibnamefont {Doyle}},\ and\ \bibinfo {author}
  {\bibfnamefont {J.}~\bibnamefont {Ye}},\ }\bibfield  {title} {\bibinfo
  {title} {{3D Magneto-Optical Trap of Yttrium Monoxide}},\ }\href
  {https://doi.org/10.1103/PhysRevLett.121.213201} {\bibfield  {journal}
  {\bibinfo  {journal} {Physical Review Letters}\ }\textbf {\bibinfo {volume}
  {121}},\ \bibinfo {pages} {213201} (\bibinfo {year} {2018})},\ \Eprint
  {https://arxiv.org/abs/1808.01067} {arXiv:1808.01067} \BibitemShut {NoStop}%
\bibitem [{\citenamefont {Vilas}\ \emph {et~al.}(2022)\citenamefont {Vilas},
  \citenamefont {Hallas}, \citenamefont {Anderegg}, \citenamefont {Robichaud},
  \citenamefont {Winnicki}, \citenamefont {Mitra},\ and\ \citenamefont
  {Doyle}}]{Vilas2021}%
  \BibitemOpen
  \bibfield  {author} {\bibinfo {author} {\bibfnamefont {N.~B.}\ \bibnamefont
  {Vilas}}, \bibinfo {author} {\bibfnamefont {C.}~\bibnamefont {Hallas}},
  \bibinfo {author} {\bibfnamefont {L.}~\bibnamefont {Anderegg}}, \bibinfo
  {author} {\bibfnamefont {P.}~\bibnamefont {Robichaud}}, \bibinfo {author}
  {\bibfnamefont {A.}~\bibnamefont {Winnicki}}, \bibinfo {author}
  {\bibfnamefont {D.}~\bibnamefont {Mitra}},\ and\ \bibinfo {author}
  {\bibfnamefont {J.~M.}\ \bibnamefont {Doyle}},\ }\bibfield  {title} {\bibinfo
  {title} {{Magneto-optical trapping and sub-Doppler cooling of a polyatomic
  molecule}},\ }\href {https://doi.org/10.1038/s41586-022-04620-5} {\bibfield
  {journal} {\bibinfo  {journal} {Nature}\ }\textbf {\bibinfo {volume} {606}},\
  \bibinfo {pages} {70} (\bibinfo {year} {2022})}\BibitemShut {NoStop}%
\bibitem [{\citenamefont {McCarron}\ \emph {et~al.}(2018)\citenamefont
  {McCarron}, \citenamefont {Steinecker}, \citenamefont {Zhu},\ and\
  \citenamefont {Demille}}]{McCarron2018}%
  \BibitemOpen
  \bibfield  {author} {\bibinfo {author} {\bibfnamefont {D.~J.}\ \bibnamefont
  {McCarron}}, \bibinfo {author} {\bibfnamefont {M.~H.}\ \bibnamefont
  {Steinecker}}, \bibinfo {author} {\bibfnamefont {Y.}~\bibnamefont {Zhu}},\
  and\ \bibinfo {author} {\bibfnamefont {D.}~\bibnamefont {Demille}},\
  }\bibfield  {title} {\bibinfo {title} {{Magnetic Trapping of an Ultracold Gas
  of Polar Molecules}},\ }\href
  {https://doi.org/10.1103/PhysRevLett.121.013202} {\bibfield  {journal}
  {\bibinfo  {journal} {Physical Review Letters}\ }\textbf {\bibinfo {volume}
  {121}},\ \bibinfo {pages} {13202} (\bibinfo {year} {2018})}\BibitemShut
  {NoStop}%
\bibitem [{\citenamefont {Williams}\ \emph {et~al.}(2018)\citenamefont
  {Williams}, \citenamefont {Caldwell}, \citenamefont {Fitch}, \citenamefont
  {Truppe}, \citenamefont {Rodewald}, \citenamefont {Hinds}, \citenamefont
  {Sauer},\ and\ \citenamefont {Tarbutt}}]{Williams2018}%
  \BibitemOpen
  \bibfield  {author} {\bibinfo {author} {\bibfnamefont {H.~J.}\ \bibnamefont
  {Williams}}, \bibinfo {author} {\bibfnamefont {L.}~\bibnamefont {Caldwell}},
  \bibinfo {author} {\bibfnamefont {N.~J.}\ \bibnamefont {Fitch}}, \bibinfo
  {author} {\bibfnamefont {S.}~\bibnamefont {Truppe}}, \bibinfo {author}
  {\bibfnamefont {J.}~\bibnamefont {Rodewald}}, \bibinfo {author}
  {\bibfnamefont {E.~A.}\ \bibnamefont {Hinds}}, \bibinfo {author}
  {\bibfnamefont {B.~E.}\ \bibnamefont {Sauer}},\ and\ \bibinfo {author}
  {\bibfnamefont {M.~R.}\ \bibnamefont {Tarbutt}},\ }\bibfield  {title}
  {\bibinfo {title} {Magnetic trapping and coherent control of laser-cooled
  molecules},\ }\href {https://doi.org/10.1103/PhysRevLett.120.163201}
  {\bibfield  {journal} {\bibinfo  {journal} {Phys. Rev. Lett.}\ }\textbf
  {\bibinfo {volume} {120}},\ \bibinfo {pages} {163201} (\bibinfo {year}
  {2018})}\BibitemShut {NoStop}%
\bibitem [{\citenamefont {Wu}\ \emph {et~al.}(2021)\citenamefont {Wu},
  \citenamefont {Burau}, \citenamefont {Mehling}, \citenamefont {Ye},\ and\
  \citenamefont {Ding}}]{Wu2021}%
  \BibitemOpen
  \bibfield  {author} {\bibinfo {author} {\bibfnamefont {Y.}~\bibnamefont
  {Wu}}, \bibinfo {author} {\bibfnamefont {J.~J.}\ \bibnamefont {Burau}},
  \bibinfo {author} {\bibfnamefont {K.}~\bibnamefont {Mehling}}, \bibinfo
  {author} {\bibfnamefont {J.}~\bibnamefont {Ye}},\ and\ \bibinfo {author}
  {\bibfnamefont {S.}~\bibnamefont {Ding}},\ }\bibfield  {title} {\bibinfo
  {title} {High phase-space density of laser-cooled molecules in an optical
  lattice},\ }\href {https://doi.org/10.1103/PhysRevLett.127.263201} {\bibfield
   {journal} {\bibinfo  {journal} {Phys. Rev. Lett.}\ }\textbf {\bibinfo
  {volume} {127}},\ \bibinfo {pages} {263201} (\bibinfo {year}
  {2021})}\BibitemShut {NoStop}%
\bibitem [{\citenamefont {Anderegg}\ \emph {et~al.}(2019)\citenamefont
  {Anderegg}, \citenamefont {Cheuk}, \citenamefont {Bao}, \citenamefont
  {Burchesky}, \citenamefont {Ketterle}, \citenamefont {Ni},\ and\
  \citenamefont {Doyle}}]{Anderegg2019}%
  \BibitemOpen
  \bibfield  {author} {\bibinfo {author} {\bibfnamefont {L.}~\bibnamefont
  {Anderegg}}, \bibinfo {author} {\bibfnamefont {L.~W.}\ \bibnamefont {Cheuk}},
  \bibinfo {author} {\bibfnamefont {Y.}~\bibnamefont {Bao}}, \bibinfo {author}
  {\bibfnamefont {S.}~\bibnamefont {Burchesky}}, \bibinfo {author}
  {\bibfnamefont {W.}~\bibnamefont {Ketterle}}, \bibinfo {author}
  {\bibfnamefont {K.-K.}\ \bibnamefont {Ni}},\ and\ \bibinfo {author}
  {\bibfnamefont {J.~M.}\ \bibnamefont {Doyle}},\ }\bibfield  {title} {\bibinfo
  {title} {{An optical tweezer array of ultracold molecules}},\ }\href
  {https://doi.org/10.1126/science.aax1265} {\bibfield  {journal} {\bibinfo
  {journal} {Science}\ }\textbf {\bibinfo {volume} {365}},\ \bibinfo {pages}
  {1156} (\bibinfo {year} {2019})}\BibitemShut {NoStop}%
\bibitem [{\citenamefont {Vilas}\ \emph {et~al.}(2024)\citenamefont {Vilas},
  \citenamefont {Robichaud}, \citenamefont {Hallas}, \citenamefont {Li},
  \citenamefont {Anderegg},\ and\ \citenamefont {Doyle}}]{Vilas2023}%
  \BibitemOpen
  \bibfield  {author} {\bibinfo {author} {\bibfnamefont {N.~B.}\ \bibnamefont
  {Vilas}}, \bibinfo {author} {\bibfnamefont {P.}~\bibnamefont {Robichaud}},
  \bibinfo {author} {\bibfnamefont {C.}~\bibnamefont {Hallas}}, \bibinfo
  {author} {\bibfnamefont {G.~K.}\ \bibnamefont {Li}}, \bibinfo {author}
  {\bibfnamefont {L.}~\bibnamefont {Anderegg}},\ and\ \bibinfo {author}
  {\bibfnamefont {J.~M.}\ \bibnamefont {Doyle}},\ }\bibfield  {title} {\bibinfo
  {title} {An optical tweezer array of ultracold polyatomic molecules},\ }\href
  {https://doi.org/10.1038/s41586-024-07199-1} {\bibfield  {journal} {\bibinfo
  {journal} {Nature}\ }\textbf {\bibinfo {volume} {628}},\ \bibinfo {pages}
  {282} (\bibinfo {year} {2024})}\BibitemShut {NoStop}%
\bibitem [{\citenamefont {Cheuk}\ \emph {et~al.}(2020)\citenamefont {Cheuk},
  \citenamefont {Anderegg}, \citenamefont {Bao}, \citenamefont {Burchesky},
  \citenamefont {Yu}, \citenamefont {Ketterle}, \citenamefont {Ni},\ and\
  \citenamefont {Doyle}}]{Cheuk2020}%
  \BibitemOpen
  \bibfield  {author} {\bibinfo {author} {\bibfnamefont {L.~W.}\ \bibnamefont
  {Cheuk}}, \bibinfo {author} {\bibfnamefont {L.}~\bibnamefont {Anderegg}},
  \bibinfo {author} {\bibfnamefont {Y.}~\bibnamefont {Bao}}, \bibinfo {author}
  {\bibfnamefont {S.}~\bibnamefont {Burchesky}}, \bibinfo {author}
  {\bibfnamefont {S.~S.}\ \bibnamefont {Yu}}, \bibinfo {author} {\bibfnamefont
  {W.}~\bibnamefont {Ketterle}}, \bibinfo {author} {\bibfnamefont {K.-K.}\
  \bibnamefont {Ni}},\ and\ \bibinfo {author} {\bibfnamefont {J.~M.}\
  \bibnamefont {Doyle}},\ }\bibfield  {title} {\bibinfo {title} {{Observation
  of Collisions between Two Ultracold Ground-State CaF Molecules}},\ }\href
  {https://doi.org/10.1103/PhysRevLett.125.043401} {\bibfield  {journal}
  {\bibinfo  {journal} {Phys. Rev. Lett.}\ }\textbf {\bibinfo {volume} {125}},\
  \bibinfo {pages} {43401} (\bibinfo {year} {2020})}\BibitemShut {NoStop}%
\bibitem [{\citenamefont {Skripnikov}\ \emph {et~al.}(2014)\citenamefont
  {Skripnikov}, \citenamefont {Petrov}, \citenamefont {Titov},\ and\
  \citenamefont {Flambaum}}]{Skripnikov2014}%
  \BibitemOpen
  \bibfield  {author} {\bibinfo {author} {\bibfnamefont {L.~V.}\ \bibnamefont
  {Skripnikov}}, \bibinfo {author} {\bibfnamefont {A.~N.}\ \bibnamefont
  {Petrov}}, \bibinfo {author} {\bibfnamefont {A.~V.}\ \bibnamefont {Titov}},\
  and\ \bibinfo {author} {\bibfnamefont {V.~V.}\ \bibnamefont {Flambaum}},\
  }\bibfield  {title} {\bibinfo {title} {$cp$-violating effect of the th
  nuclear magnetic quadrupole moment: Accurate many-body study of tho},\ }\href
  {https://doi.org/10.1103/PhysRevLett.113.263006} {\bibfield  {journal}
  {\bibinfo  {journal} {Phys. Rev. Lett.}\ }\textbf {\bibinfo {volume} {113}},\
  \bibinfo {pages} {263006} (\bibinfo {year} {2014})}\BibitemShut {NoStop}%
\bibitem [{\citenamefont {Lackenby}\ and\ \citenamefont
  {Flambaum}(2018)}]{Lackenby2018}%
  \BibitemOpen
  \bibfield  {author} {\bibinfo {author} {\bibfnamefont {B.~G.~C.}\
  \bibnamefont {Lackenby}}\ and\ \bibinfo {author} {\bibfnamefont {V.~V.}\
  \bibnamefont {Flambaum}},\ }\bibfield  {title} {\bibinfo {title} {Time
  reversal violating magnetic quadrupole moment in heavy deformed nuclei},\
  }\href {https://doi.org/10.1103/PhysRevD.98.115019} {\bibfield  {journal}
  {\bibinfo  {journal} {Phys. Rev. D}\ }\textbf {\bibinfo {volume} {98}},\
  \bibinfo {pages} {115019} (\bibinfo {year} {2018})}\BibitemShut {NoStop}%
\bibitem [{\citenamefont {Arrowsmith-Kron}\ \emph {et~al.}(2023)\citenamefont
  {Arrowsmith-Kron}, \citenamefont {Athanasakis-Kaklamanakis}, \citenamefont
  {Au}, \citenamefont {Ballof}, \citenamefont {Berger}, \citenamefont
  {Borschevsky}, \citenamefont {Breier}, \citenamefont {Buchinger},
  \citenamefont {Budker}, \citenamefont {Caldwell}, \citenamefont {Charles},
  \citenamefont {Dattani}, \citenamefont {de~Groote}, \citenamefont {DeMille},
  \citenamefont {Dickel}, \citenamefont {Dobaczewski}, \citenamefont
  {Düllmann}, \citenamefont {Eliav}, \citenamefont {Engel}, \citenamefont
  {Fan}, \citenamefont {Flambaum}, \citenamefont {Flanagan}, \citenamefont
  {Gaiser}, \citenamefont {Ruiz}, \citenamefont {Gaul}, \citenamefont {Giesen},
  \citenamefont {Ginges}, \citenamefont {Gottberg}, \citenamefont {Gwinner},
  \citenamefont {Heinke}, \citenamefont {Hoekstra}, \citenamefont {Holt},
  \citenamefont {Hutzler}, \citenamefont {Jayich}, \citenamefont {Karthein},
  \citenamefont {Leach}, \citenamefont {Madison}, \citenamefont
  {Malbrunot-Ettenauer}, \citenamefont {Miyagi}, \citenamefont {Moore},
  \citenamefont {Moroch}, \citenamefont {Navrátil}, \citenamefont
  {Nazarewicz}, \citenamefont {Neyens}, \citenamefont {Norrgard}, \citenamefont
  {Nusgart}, \citenamefont {Pašteka}, \citenamefont {Petrov}, \citenamefont
  {Plass}, \citenamefont {Ready}, \citenamefont {Reiter}, \citenamefont
  {Reponen}, \citenamefont {Rothe}, \citenamefont {Safronova}, \citenamefont
  {Scheidenberger}, \citenamefont {Shindler}, \citenamefont {Singh},
  \citenamefont {Skripnikov}, \citenamefont {Titov}, \citenamefont {Udrescu},
  \citenamefont {Wilkins},\ and\ \citenamefont {Yang}}]{ArrowsmithKron2023}%
  \BibitemOpen
  \bibfield  {author} {\bibinfo {author} {\bibfnamefont {G.}~\bibnamefont
  {Arrowsmith-Kron}}, \bibinfo {author} {\bibfnamefont {M.}~\bibnamefont
  {Athanasakis-Kaklamanakis}}, \bibinfo {author} {\bibfnamefont
  {M.}~\bibnamefont {Au}}, \bibinfo {author} {\bibfnamefont {J.}~\bibnamefont
  {Ballof}}, \bibinfo {author} {\bibfnamefont {R.}~\bibnamefont {Berger}},
  \bibinfo {author} {\bibfnamefont {A.}~\bibnamefont {Borschevsky}}, \bibinfo
  {author} {\bibfnamefont {A.~A.}\ \bibnamefont {Breier}}, \bibinfo {author}
  {\bibfnamefont {F.}~\bibnamefont {Buchinger}}, \bibinfo {author}
  {\bibfnamefont {D.}~\bibnamefont {Budker}}, \bibinfo {author} {\bibfnamefont
  {L.}~\bibnamefont {Caldwell}}, \bibinfo {author} {\bibfnamefont
  {C.}~\bibnamefont {Charles}}, \bibinfo {author} {\bibfnamefont
  {N.}~\bibnamefont {Dattani}}, \bibinfo {author} {\bibfnamefont {R.~P.}\
  \bibnamefont {de~Groote}}, \bibinfo {author} {\bibfnamefont {D.}~\bibnamefont
  {DeMille}}, \bibinfo {author} {\bibfnamefont {T.}~\bibnamefont {Dickel}},
  \bibinfo {author} {\bibfnamefont {J.}~\bibnamefont {Dobaczewski}}, \bibinfo
  {author} {\bibfnamefont {C.~E.}\ \bibnamefont {Düllmann}}, \bibinfo {author}
  {\bibfnamefont {E.}~\bibnamefont {Eliav}}, \bibinfo {author} {\bibfnamefont
  {J.}~\bibnamefont {Engel}}, \bibinfo {author} {\bibfnamefont
  {M.}~\bibnamefont {Fan}}, \bibinfo {author} {\bibfnamefont {V.}~\bibnamefont
  {Flambaum}}, \bibinfo {author} {\bibfnamefont {K.~T.}\ \bibnamefont
  {Flanagan}}, \bibinfo {author} {\bibfnamefont {A.}~\bibnamefont {Gaiser}},
  \bibinfo {author} {\bibfnamefont {R.~G.}\ \bibnamefont {Ruiz}}, \bibinfo
  {author} {\bibfnamefont {K.}~\bibnamefont {Gaul}}, \bibinfo {author}
  {\bibfnamefont {T.~F.}\ \bibnamefont {Giesen}}, \bibinfo {author}
  {\bibfnamefont {J.}~\bibnamefont {Ginges}}, \bibinfo {author} {\bibfnamefont
  {A.}~\bibnamefont {Gottberg}}, \bibinfo {author} {\bibfnamefont
  {G.}~\bibnamefont {Gwinner}}, \bibinfo {author} {\bibfnamefont
  {R.}~\bibnamefont {Heinke}}, \bibinfo {author} {\bibfnamefont
  {S.}~\bibnamefont {Hoekstra}}, \bibinfo {author} {\bibfnamefont {J.~D.}\
  \bibnamefont {Holt}}, \bibinfo {author} {\bibfnamefont {N.~R.}\ \bibnamefont
  {Hutzler}}, \bibinfo {author} {\bibfnamefont {A.}~\bibnamefont {Jayich}},
  \bibinfo {author} {\bibfnamefont {J.}~\bibnamefont {Karthein}}, \bibinfo
  {author} {\bibfnamefont {K.~G.}\ \bibnamefont {Leach}}, \bibinfo {author}
  {\bibfnamefont {K.}~\bibnamefont {Madison}}, \bibinfo {author} {\bibfnamefont
  {S.}~\bibnamefont {Malbrunot-Ettenauer}}, \bibinfo {author} {\bibfnamefont
  {T.}~\bibnamefont {Miyagi}}, \bibinfo {author} {\bibfnamefont {I.~D.}\
  \bibnamefont {Moore}}, \bibinfo {author} {\bibfnamefont {S.}~\bibnamefont
  {Moroch}}, \bibinfo {author} {\bibfnamefont {P.}~\bibnamefont {Navrátil}},
  \bibinfo {author} {\bibfnamefont {W.}~\bibnamefont {Nazarewicz}}, \bibinfo
  {author} {\bibfnamefont {G.}~\bibnamefont {Neyens}}, \bibinfo {author}
  {\bibfnamefont {E.}~\bibnamefont {Norrgard}}, \bibinfo {author}
  {\bibfnamefont {N.}~\bibnamefont {Nusgart}}, \bibinfo {author} {\bibfnamefont
  {L.~F.}\ \bibnamefont {Pašteka}}, \bibinfo {author} {\bibfnamefont {A.~N.}\
  \bibnamefont {Petrov}}, \bibinfo {author} {\bibfnamefont {W.}~\bibnamefont
  {Plass}}, \bibinfo {author} {\bibfnamefont {R.~A.}\ \bibnamefont {Ready}},
  \bibinfo {author} {\bibfnamefont {M.~P.}\ \bibnamefont {Reiter}}, \bibinfo
  {author} {\bibfnamefont {M.}~\bibnamefont {Reponen}}, \bibinfo {author}
  {\bibfnamefont {S.}~\bibnamefont {Rothe}}, \bibinfo {author} {\bibfnamefont
  {M.}~\bibnamefont {Safronova}}, \bibinfo {author} {\bibfnamefont
  {C.}~\bibnamefont {Scheidenberger}}, \bibinfo {author} {\bibfnamefont
  {A.}~\bibnamefont {Shindler}}, \bibinfo {author} {\bibfnamefont {J.~T.}\
  \bibnamefont {Singh}}, \bibinfo {author} {\bibfnamefont {L.~V.}\ \bibnamefont
  {Skripnikov}}, \bibinfo {author} {\bibfnamefont {A.~V.}\ \bibnamefont
  {Titov}}, \bibinfo {author} {\bibfnamefont {S.-M.}\ \bibnamefont {Udrescu}},
  \bibinfo {author} {\bibfnamefont {S.~G.}\ \bibnamefont {Wilkins}},\ and\
  \bibinfo {author} {\bibfnamefont {X.}~\bibnamefont {Yang}},\ }\href@noop {}
  {\bibinfo {title} {Opportunities for fundamental physics research with
  radioactive molecules}} (\bibinfo {year} {2023}),\ \Eprint
  {https://arxiv.org/abs/2302.02165} {arXiv:2302.02165 [nucl-ex]} \BibitemShut
  {NoStop}%
\bibitem [{\citenamefont {Tiberi}\ \emph {et~al.}(2024)\citenamefont {Tiberi},
  \citenamefont {Borkowski}, \citenamefont {Iritani}, \citenamefont
  {Moszynski},\ and\ \citenamefont {Zelevinsky}}]{Tiberi2024}%
  \BibitemOpen
  \bibfield  {author} {\bibinfo {author} {\bibfnamefont {E.}~\bibnamefont
  {Tiberi}}, \bibinfo {author} {\bibfnamefont {M.}~\bibnamefont {Borkowski}},
  \bibinfo {author} {\bibfnamefont {B.}~\bibnamefont {Iritani}}, \bibinfo
  {author} {\bibfnamefont {R.}~\bibnamefont {Moszynski}},\ and\ \bibinfo
  {author} {\bibfnamefont {T.}~\bibnamefont {Zelevinsky}},\ }\href@noop {}
  {\bibinfo {title} {Searching for new fundamental interactions via isotopic
  shifts in molecular lattice clocks}} (\bibinfo {year} {2024}),\ \Eprint
  {https://arxiv.org/abs/2403.07097} {arXiv:2403.07097 [physics.atom-ph]}
  \BibitemShut {NoStop}%
\bibitem [{\citenamefont {DeMille}\ \emph {et~al.}(2008)\citenamefont
  {DeMille}, \citenamefont {Cahn}, \citenamefont {Murphree}, \citenamefont
  {Rahmlow},\ and\ \citenamefont {Kozlov}}]{Demille2008}%
  \BibitemOpen
  \bibfield  {author} {\bibinfo {author} {\bibfnamefont {D.}~\bibnamefont
  {DeMille}}, \bibinfo {author} {\bibfnamefont {S.~B.}\ \bibnamefont {Cahn}},
  \bibinfo {author} {\bibfnamefont {D.}~\bibnamefont {Murphree}}, \bibinfo
  {author} {\bibfnamefont {D.~A.}\ \bibnamefont {Rahmlow}},\ and\ \bibinfo
  {author} {\bibfnamefont {M.~G.}\ \bibnamefont {Kozlov}},\ }\bibfield  {title}
  {\bibinfo {title} {{Using Molecules to Measure Nuclear Spin-Dependent Parity
  Violation}},\ }\href {https://doi.org/10.1103/PhysRevLett.100.023003}
  {\bibfield  {journal} {\bibinfo  {journal} {Physical Review Letters}\
  }\textbf {\bibinfo {volume} {100}},\ \bibinfo {pages} {023003} (\bibinfo
  {year} {2008})}\BibitemShut {NoStop}%
\bibitem [{\citenamefont {Altuntas}\ \emph {et~al.}(2018)\citenamefont
  {Altuntas}, \citenamefont {Ammon}, \citenamefont {Cahn},\ and\ \citenamefont
  {DeMille}}]{Altuntas2018}%
  \BibitemOpen
  \bibfield  {author} {\bibinfo {author} {\bibfnamefont {E.}~\bibnamefont
  {Altuntas}}, \bibinfo {author} {\bibfnamefont {J.}~\bibnamefont {Ammon}},
  \bibinfo {author} {\bibfnamefont {S.~B.}\ \bibnamefont {Cahn}},\ and\
  \bibinfo {author} {\bibfnamefont {D.}~\bibnamefont {DeMille}},\ }\bibfield
  {title} {\bibinfo {title} {{Demonstration of a Sensitive Method to Measure
  Nuclear-Spin-Dependent Parity Violation}},\ }\href
  {https://doi.org/10.1103/PhysRevLett.120.142501} {\bibfield  {journal}
  {\bibinfo  {journal} {Phys. Rev. Lett.}\ }\textbf {\bibinfo {volume} {120}},\
  \bibinfo {pages} {142501} (\bibinfo {year} {2018})}\BibitemShut {NoStop}%
\bibitem [{\citenamefont {Antypas}\ \emph {et~al.}(2019)\citenamefont
  {Antypas}, \citenamefont {Fabricant}, \citenamefont {Stalnaker},
  \citenamefont {Tsigutkin}, \citenamefont {Flambaum},\ and\ \citenamefont
  {Budker}}]{Antypas2019}%
  \BibitemOpen
  \bibfield  {author} {\bibinfo {author} {\bibfnamefont {D.}~\bibnamefont
  {Antypas}}, \bibinfo {author} {\bibfnamefont {A.}~\bibnamefont {Fabricant}},
  \bibinfo {author} {\bibfnamefont {J.~E.}\ \bibnamefont {Stalnaker}}, \bibinfo
  {author} {\bibfnamefont {K.}~\bibnamefont {Tsigutkin}}, \bibinfo {author}
  {\bibfnamefont {V.~V.}\ \bibnamefont {Flambaum}},\ and\ \bibinfo {author}
  {\bibfnamefont {D.}~\bibnamefont {Budker}},\ }\bibfield  {title} {\bibinfo
  {title} {Isotopic variation of parity violation in atomic ytterbium},\ }\href
  {https://doi.org/10.1038/s41567-018-0312-8} {\bibfield  {journal} {\bibinfo
  {journal} {Nature Physics}\ }\textbf {\bibinfo {volume} {15}},\ \bibinfo
  {pages} {120} (\bibinfo {year} {2019})}\BibitemShut {NoStop}%
\bibitem [{\citenamefont {Hao}\ \emph {et~al.}(2020)\citenamefont {Hao},
  \citenamefont {Navr\'atil}, \citenamefont {Norrgard}, \citenamefont
  {Ilia\ifmmode~\check{s}\else \v{s}\fi{}}, \citenamefont {Eliav},
  \citenamefont {Timmermans}, \citenamefont {Flambaum},\ and\ \citenamefont
  {Borschevsky}}]{Hao2020}%
  \BibitemOpen
  \bibfield  {author} {\bibinfo {author} {\bibfnamefont {Y.}~\bibnamefont
  {Hao}}, \bibinfo {author} {\bibfnamefont {P.}~\bibnamefont {Navr\'atil}},
  \bibinfo {author} {\bibfnamefont {E.~B.}\ \bibnamefont {Norrgard}}, \bibinfo
  {author} {\bibfnamefont {M.}~\bibnamefont {Ilia\ifmmode~\check{s}\else
  \v{s}\fi{}}}, \bibinfo {author} {\bibfnamefont {E.}~\bibnamefont {Eliav}},
  \bibinfo {author} {\bibfnamefont {R.~G.~E.}\ \bibnamefont {Timmermans}},
  \bibinfo {author} {\bibfnamefont {V.~V.}\ \bibnamefont {Flambaum}},\ and\
  \bibinfo {author} {\bibfnamefont {A.}~\bibnamefont {Borschevsky}},\
  }\bibfield  {title} {\bibinfo {title} {Nuclear spin-dependent
  parity-violating effects in light polyatomic molecules},\ }\href
  {https://doi.org/10.1103/PhysRevA.102.052828} {\bibfield  {journal} {\bibinfo
   {journal} {Phys. Rev. A}\ }\textbf {\bibinfo {volume} {102}},\ \bibinfo
  {pages} {052828} (\bibinfo {year} {2020})}\BibitemShut {NoStop}%
\bibitem [{\citenamefont {Tomza}(2015)}]{Tomza2015}%
  \BibitemOpen
  \bibfield  {author} {\bibinfo {author} {\bibfnamefont {M.}~\bibnamefont
  {Tomza}},\ }\bibfield  {title} {\bibinfo {title} {Energetics and control of
  ultracold isotope-exchange reactions between heteronuclear dimers in external
  fields},\ }\href {https://doi.org/10.1103/PhysRevLett.115.063201} {\bibfield
  {journal} {\bibinfo  {journal} {Phys. Rev. Lett.}\ }\textbf {\bibinfo
  {volume} {115}},\ \bibinfo {pages} {063201} (\bibinfo {year}
  {2015})}\BibitemShut {NoStop}%
\bibitem [{\citenamefont {Fleurat-Lessard}\ \emph {et~al.}(2003)\citenamefont
  {Fleurat-Lessard}, \citenamefont {Grebenshchikov}, \citenamefont {Schinke},
  \citenamefont {Janssen},\ and\ \citenamefont {Krankowsky}}]{Fleurat2003}%
  \BibitemOpen
  \bibfield  {author} {\bibinfo {author} {\bibfnamefont {P.}~\bibnamefont
  {Fleurat-Lessard}}, \bibinfo {author} {\bibfnamefont {S.~Y.}\ \bibnamefont
  {Grebenshchikov}}, \bibinfo {author} {\bibfnamefont {R.}~\bibnamefont
  {Schinke}}, \bibinfo {author} {\bibfnamefont {C.}~\bibnamefont {Janssen}},\
  and\ \bibinfo {author} {\bibfnamefont {D.}~\bibnamefont {Krankowsky}},\
  }\bibfield  {title} {\bibinfo {title} {{Isotope dependence of the O+O2
  exchange reaction: Experiment and theory}},\ }\href
  {https://doi.org/10.1063/1.1595091} {\bibfield  {journal} {\bibinfo
  {journal} {The Journal of Chemical Physics}\ }\textbf {\bibinfo {volume}
  {119}},\ \bibinfo {pages} {4700} (\bibinfo {year} {2003})}\BibitemShut
  {NoStop}%
\bibitem [{\citenamefont {Brenninkmeijer}\ \emph {et~al.}(2003)\citenamefont
  {Brenninkmeijer}, \citenamefont {Janssen}, \citenamefont {Kaiser},
  \citenamefont {R{\"o}ckmann}, \citenamefont {Rhee},\ and\ \citenamefont
  {Assonov}}]{Brenninkmeijer2003}%
  \BibitemOpen
  \bibfield  {author} {\bibinfo {author} {\bibfnamefont {C.~A.~M.}\
  \bibnamefont {Brenninkmeijer}}, \bibinfo {author} {\bibfnamefont
  {C.}~\bibnamefont {Janssen}}, \bibinfo {author} {\bibfnamefont
  {J.}~\bibnamefont {Kaiser}}, \bibinfo {author} {\bibfnamefont
  {T.}~\bibnamefont {R{\"o}ckmann}}, \bibinfo {author} {\bibfnamefont {T.~S.}\
  \bibnamefont {Rhee}},\ and\ \bibinfo {author} {\bibfnamefont {S.~S.}\
  \bibnamefont {Assonov}},\ }\bibfield  {title} {\bibinfo {title} {Isotope
  effects in the chemistry of atmospheric trace compounds},\ }\href
  {https://doi.org/10.1021/cr020644k} {\bibfield  {journal} {\bibinfo
  {journal} {Chemical Reviews}\ }\textbf {\bibinfo {volume} {103}},\ \bibinfo
  {pages} {5125} (\bibinfo {year} {2003})}\BibitemShut {NoStop}%
\bibitem [{\citenamefont {Fleischer}\ \emph {et~al.}(2006)\citenamefont
  {Fleischer}, \citenamefont {Averbukh},\ and\ \citenamefont
  {Prior}}]{Fleischer2006}%
  \BibitemOpen
  \bibfield  {author} {\bibinfo {author} {\bibfnamefont {S.}~\bibnamefont
  {Fleischer}}, \bibinfo {author} {\bibfnamefont {I.~S.}\ \bibnamefont
  {Averbukh}},\ and\ \bibinfo {author} {\bibfnamefont {Y.}~\bibnamefont
  {Prior}},\ }\bibfield  {title} {\bibinfo {title} {Isotope-selective laser
  molecular alignment},\ }\href {https://doi.org/10.1103/PhysRevA.74.041403}
  {\bibfield  {journal} {\bibinfo  {journal} {Phys. Rev. A}\ }\textbf {\bibinfo
  {volume} {74}},\ \bibinfo {pages} {041403} (\bibinfo {year}
  {2006})}\BibitemShut {NoStop}%
\bibitem [{\citenamefont {{Visser, R.}}\ \emph {et~al.}(2009)\citenamefont
  {{Visser, R.}}, \citenamefont {{van Dishoeck, E. F.}},\ and\ \citenamefont
  {{Black, J. H.}}}]{Visser2009}%
  \BibitemOpen
  \bibfield  {author} {\bibinfo {author} {\bibnamefont {{Visser, R.}}},
  \bibinfo {author} {\bibnamefont {{van Dishoeck, E. F.}}},\ and\ \bibinfo
  {author} {\bibnamefont {{Black, J. H.}}},\ }\bibfield  {title} {\bibinfo
  {title} {The photodissociation and chemistry of co isotopologues:
  applications to interstellar clouds and circumstellar disks*},\ }\href
  {https://doi.org/10.1051/0004-6361/200912129} {\bibfield  {journal} {\bibinfo
   {journal} {Astronomy \& Astrophysics}\ }\textbf {\bibinfo {volume} {503}},\
  \bibinfo {pages} {323} (\bibinfo {year} {2009})}\BibitemShut {NoStop}%
\bibitem [{\citenamefont {Zhdanovich}\ \emph {et~al.}(2012)\citenamefont
  {Zhdanovich}, \citenamefont {Bloomquist}, \citenamefont {Flo\ss{}},
  \citenamefont {Averbukh}, \citenamefont {Hepburn},\ and\ \citenamefont
  {Milner}}]{Zhdanovich2012}%
  \BibitemOpen
  \bibfield  {author} {\bibinfo {author} {\bibfnamefont {S.}~\bibnamefont
  {Zhdanovich}}, \bibinfo {author} {\bibfnamefont {C.}~\bibnamefont
  {Bloomquist}}, \bibinfo {author} {\bibfnamefont {J.}~\bibnamefont
  {Flo\ss{}}}, \bibinfo {author} {\bibfnamefont {I.~S.}\ \bibnamefont
  {Averbukh}}, \bibinfo {author} {\bibfnamefont {J.~W.}\ \bibnamefont
  {Hepburn}},\ and\ \bibinfo {author} {\bibfnamefont {V.}~\bibnamefont
  {Milner}},\ }\bibfield  {title} {\bibinfo {title} {Quantum resonances in
  selective rotational excitation of molecules with a sequence of ultrashort
  laser pulses},\ }\href {https://doi.org/10.1103/PhysRevLett.109.043003}
  {\bibfield  {journal} {\bibinfo  {journal} {Phys. Rev. Lett.}\ }\textbf
  {\bibinfo {volume} {109}},\ \bibinfo {pages} {043003} (\bibinfo {year}
  {2012})}\BibitemShut {NoStop}%
\bibitem [{\citenamefont {Griffith}(2018)}]{Griffith2018}%
  \BibitemOpen
  \bibfield  {author} {\bibinfo {author} {\bibfnamefont {D.~W.~T.}\
  \bibnamefont {Griffith}},\ }\bibfield  {title} {\bibinfo {title} {Calibration
  of isotopologue-specific optical trace gas analysers: a practical guide},\
  }\href {https://doi.org/10.5194/amt-11-6189-2018} {\bibfield  {journal}
  {\bibinfo  {journal} {Atmospheric Measurement Techniques}\ }\textbf {\bibinfo
  {volume} {11}},\ \bibinfo {pages} {6189} (\bibinfo {year}
  {2018})}\BibitemShut {NoStop}%
\bibitem [{\citenamefont {Kami{\'n}ski}\ \emph {et~al.}(2018)\citenamefont
  {Kami{\'n}ski}, \citenamefont {Tylenda}, \citenamefont {Menten},
  \citenamefont {Karakas}, \citenamefont {Winters}, \citenamefont {Breier},
  \citenamefont {Wong}, \citenamefont {Giesen},\ and\ \citenamefont
  {Patel}}]{Kaminsky2018}%
  \BibitemOpen
  \bibfield  {author} {\bibinfo {author} {\bibfnamefont {T.}~\bibnamefont
  {Kami{\'n}ski}}, \bibinfo {author} {\bibfnamefont {R.}~\bibnamefont
  {Tylenda}}, \bibinfo {author} {\bibfnamefont {K.~M.}\ \bibnamefont {Menten}},
  \bibinfo {author} {\bibfnamefont {A.}~\bibnamefont {Karakas}}, \bibinfo
  {author} {\bibfnamefont {J.~M.}\ \bibnamefont {Winters}}, \bibinfo {author}
  {\bibfnamefont {A.~A.}\ \bibnamefont {Breier}}, \bibinfo {author}
  {\bibfnamefont {K.~T.}\ \bibnamefont {Wong}}, \bibinfo {author}
  {\bibfnamefont {T.~F.}\ \bibnamefont {Giesen}},\ and\ \bibinfo {author}
  {\bibfnamefont {N.~A.}\ \bibnamefont {Patel}},\ }\bibfield  {title} {\bibinfo
  {title} {Astronomical detection of radioactive molecule 26alf in the remnant
  of an ancient explosion},\ }\href {https://doi.org/10.1038/s41550-018-0541-x}
  {\bibfield  {journal} {\bibinfo  {journal} {Nature Astronomy}\ }\textbf
  {\bibinfo {volume} {2}},\ \bibinfo {pages} {778} (\bibinfo {year}
  {2018})}\BibitemShut {NoStop}%
\bibitem [{\citenamefont {Albrecht}\ \emph {et~al.}(2020)\citenamefont
  {Albrecht}, \citenamefont {Scharwaechter}, \citenamefont {Sixt},
  \citenamefont {Hofer},\ and\ \citenamefont {Langen}}]{Albrecht2020}%
  \BibitemOpen
  \bibfield  {author} {\bibinfo {author} {\bibfnamefont {R.}~\bibnamefont
  {Albrecht}}, \bibinfo {author} {\bibfnamefont {M.}~\bibnamefont
  {Scharwaechter}}, \bibinfo {author} {\bibfnamefont {T.}~\bibnamefont {Sixt}},
  \bibinfo {author} {\bibfnamefont {L.}~\bibnamefont {Hofer}},\ and\ \bibinfo
  {author} {\bibfnamefont {T.}~\bibnamefont {Langen}},\ }\bibfield  {title}
  {\bibinfo {title} {Buffer-gas cooling, high-resolution spectroscopy, and
  optical cycling of barium monofluoride molecules},\ }\href
  {https://doi.org/10.1103/PhysRevA.101.013413} {\bibfield  {journal} {\bibinfo
   {journal} {Phys. Rev. A}\ }\textbf {\bibinfo {volume} {101}},\ \bibinfo
  {pages} {013413} (\bibinfo {year} {2020})}\BibitemShut {NoStop}%
\bibitem [{\citenamefont {Rockenh\"auser}\ \emph {et~al.}(2023)\citenamefont
  {Rockenh\"auser}, \citenamefont {Kogel}, \citenamefont {Pultinevicius},\ and\
  \citenamefont {Langen}}]{Rockenhaeuser2023}%
  \BibitemOpen
  \bibfield  {author} {\bibinfo {author} {\bibfnamefont {M.}~\bibnamefont
  {Rockenh\"auser}}, \bibinfo {author} {\bibfnamefont {F.}~\bibnamefont
  {Kogel}}, \bibinfo {author} {\bibfnamefont {E.}~\bibnamefont
  {Pultinevicius}},\ and\ \bibinfo {author} {\bibfnamefont {T.}~\bibnamefont
  {Langen}},\ }\bibfield  {title} {\bibinfo {title} {Absorption spectroscopy
  for laser cooling and high-fidelity detection of barium monofluoride
  molecules},\ }\href {https://doi.org/10.1103/PhysRevA.108.062812} {\bibfield
  {journal} {\bibinfo  {journal} {Phys. Rev. A}\ }\textbf {\bibinfo {volume}
  {108}},\ \bibinfo {pages} {062812} (\bibinfo {year} {2023})}\BibitemShut
  {NoStop}%
\bibitem [{\citenamefont {Rockenh\"auser}\ \emph {et~al.}(2024)\citenamefont
  {Rockenh\"auser}, \citenamefont {Kogel}, \citenamefont {Garg}, \citenamefont
  {Morales},\ and\ \citenamefont {Langen}}]{Rockenhaeuser2024}%
  \BibitemOpen
  \bibfield  {author} {\bibinfo {author} {\bibfnamefont {M.}~\bibnamefont
  {Rockenh\"auser}}, \bibinfo {author} {\bibfnamefont {F.}~\bibnamefont
  {Kogel}}, \bibinfo {author} {\bibfnamefont {T.}~\bibnamefont {Garg}},
  \bibinfo {author} {\bibfnamefont {S.}~\bibnamefont {Morales}},\ and\ \bibinfo
  {author} {\bibfnamefont {T.}~\bibnamefont {Langen}},\ }\href@noop {}
  {\bibinfo {title} {Laser cooling of barium monofluoride molecules using
  optimized spectra}} (\bibinfo {year} {2024}),\ \Eprint
  {https://arxiv.org/abs/2405.09427} {arXiv:2405.09427} \BibitemShut {NoStop}%
\bibitem [{\citenamefont {Bu}\ \emph {et~al.}(2017)\citenamefont {Bu},
  \citenamefont {Chen}, \citenamefont {Lv},\ and\ \citenamefont
  {Yan}}]{Bu2017}%
  \BibitemOpen
  \bibfield  {author} {\bibinfo {author} {\bibfnamefont {W.}~\bibnamefont
  {Bu}}, \bibinfo {author} {\bibfnamefont {T.}~\bibnamefont {Chen}}, \bibinfo
  {author} {\bibfnamefont {G.}~\bibnamefont {Lv}},\ and\ \bibinfo {author}
  {\bibfnamefont {B.}~\bibnamefont {Yan}},\ }\bibfield  {title} {\bibinfo
  {title} {{Cold collision and high-resolution spectroscopy of
  buffer-gas-cooled BaF molecules}},\ }\href
  {https://doi.org/10.1103/PhysRevA.95.032701} {\bibfield  {journal} {\bibinfo
  {journal} {Physical Review A}\ }\textbf {\bibinfo {volume} {95}},\ \bibinfo
  {pages} {1} (\bibinfo {year} {2017})},\ \Eprint
  {https://arxiv.org/abs/1612.07449} {arXiv:1612.07449} \BibitemShut {NoStop}%
\bibitem [{\citenamefont {Aggarwal}\ \emph {et~al.}(2018)\citenamefont
  {Aggarwal}, \citenamefont {Bethlem}, \citenamefont {Borschevsky},
  \citenamefont {Denis}, \citenamefont {Esajas}, \citenamefont {Haase},
  \citenamefont {Hao}, \citenamefont {Hoekstra}, \citenamefont {Jungmann},
  \citenamefont {Meijknecht}, \citenamefont {Mooij}, \citenamefont
  {Timmermans}, \citenamefont {Ubachs}, \citenamefont {Willmann},\ and\
  \citenamefont {Zapara}}]{Aggarwal2018}%
  \BibitemOpen
  \bibfield  {author} {\bibinfo {author} {\bibfnamefont {P.}~\bibnamefont
  {Aggarwal}}, \bibinfo {author} {\bibfnamefont {H.~L.}\ \bibnamefont
  {Bethlem}}, \bibinfo {author} {\bibfnamefont {A.}~\bibnamefont
  {Borschevsky}}, \bibinfo {author} {\bibfnamefont {M.}~\bibnamefont {Denis}},
  \bibinfo {author} {\bibfnamefont {K.}~\bibnamefont {Esajas}}, \bibinfo
  {author} {\bibfnamefont {P.~A.}\ \bibnamefont {Haase}}, \bibinfo {author}
  {\bibfnamefont {Y.}~\bibnamefont {Hao}}, \bibinfo {author} {\bibfnamefont
  {S.}~\bibnamefont {Hoekstra}}, \bibinfo {author} {\bibfnamefont
  {K.}~\bibnamefont {Jungmann}}, \bibinfo {author} {\bibfnamefont {T.~B.}\
  \bibnamefont {Meijknecht}}, \bibinfo {author} {\bibfnamefont {M.~C.}\
  \bibnamefont {Mooij}}, \bibinfo {author} {\bibfnamefont {R.~G.}\ \bibnamefont
  {Timmermans}}, \bibinfo {author} {\bibfnamefont {W.}~\bibnamefont {Ubachs}},
  \bibinfo {author} {\bibfnamefont {L.}~\bibnamefont {Willmann}},\ and\
  \bibinfo {author} {\bibfnamefont {A.}~\bibnamefont {Zapara}},\ }\bibfield
  {title} {\bibinfo {title} {{Measuring the electric dipole moment of the
  electron in BaF}},\ }\href {https://doi.org/10.1140/epjd/e2018-90192-9}
  {\bibfield  {journal} {\bibinfo  {journal} {European Physical Journal D}\
  }\textbf {\bibinfo {volume} {72}},\ \bibinfo {pages} {197} (\bibinfo {year}
  {2018})},\ \Eprint {https://arxiv.org/abs/1804.10012} {arXiv:1804.10012}
  \BibitemShut {NoStop}%
\bibitem [{\citenamefont {Cournol}\ \emph {et~al.}(2018)\citenamefont
  {Cournol}, \citenamefont {Pillet}, \citenamefont {Lignier},\ and\
  \citenamefont {Comparat}}]{Cournol2018}%
  \BibitemOpen
  \bibfield  {author} {\bibinfo {author} {\bibfnamefont {A.}~\bibnamefont
  {Cournol}}, \bibinfo {author} {\bibfnamefont {P.}~\bibnamefont {Pillet}},
  \bibinfo {author} {\bibfnamefont {H.}~\bibnamefont {Lignier}},\ and\ \bibinfo
  {author} {\bibfnamefont {D.}~\bibnamefont {Comparat}},\ }\bibfield  {title}
  {\bibinfo {title} {{Rovibrational optical pumping of a molecular beam}},\
  }\href {https://doi.org/10.1103/PhysRevA.97.031401} {\bibfield  {journal}
  {\bibinfo  {journal} {Physical Review A}\ }\textbf {\bibinfo {volume} {97}},\
  \bibinfo {pages} {1} (\bibinfo {year} {2018})},\ \Eprint
  {https://arxiv.org/abs/1709.06797} {arXiv:1709.06797} \BibitemShut {NoStop}%
\bibitem [{\citenamefont {Hao}\ \emph {et~al.}(2019)\citenamefont {Hao},
  \citenamefont {Pa{\v{s}}teka}, \citenamefont {Visscher}, \citenamefont
  {Aggarwal}, \citenamefont {Bethlem}, \citenamefont {Boeschoten},
  \citenamefont {Borschevsky}, \citenamefont {Denis}, \citenamefont {Esajas},
  \citenamefont {Hoekstra}, \citenamefont {Jungmann}, \citenamefont {Marshall},
  \citenamefont {Meijknecht}, \citenamefont {Mooij}, \citenamefont
  {Timmermans}, \citenamefont {Touwen}, \citenamefont {Ubachs}, \citenamefont
  {Willmann}, \citenamefont {Yin},\ and\ \citenamefont {Zapara}}]{Hao2019}%
  \BibitemOpen
  \bibfield  {author} {\bibinfo {author} {\bibfnamefont {Y.}~\bibnamefont
  {Hao}}, \bibinfo {author} {\bibfnamefont {L.~F.}\ \bibnamefont
  {Pa{\v{s}}teka}}, \bibinfo {author} {\bibfnamefont {L.}~\bibnamefont
  {Visscher}}, \bibinfo {author} {\bibfnamefont {P.}~\bibnamefont {Aggarwal}},
  \bibinfo {author} {\bibfnamefont {H.~L.}\ \bibnamefont {Bethlem}}, \bibinfo
  {author} {\bibfnamefont {A.}~\bibnamefont {Boeschoten}}, \bibinfo {author}
  {\bibfnamefont {A.}~\bibnamefont {Borschevsky}}, \bibinfo {author}
  {\bibfnamefont {M.}~\bibnamefont {Denis}}, \bibinfo {author} {\bibfnamefont
  {K.}~\bibnamefont {Esajas}}, \bibinfo {author} {\bibfnamefont
  {S.}~\bibnamefont {Hoekstra}}, \bibinfo {author} {\bibfnamefont
  {K.}~\bibnamefont {Jungmann}}, \bibinfo {author} {\bibfnamefont {V.~R.}\
  \bibnamefont {Marshall}}, \bibinfo {author} {\bibfnamefont {T.~B.}\
  \bibnamefont {Meijknecht}}, \bibinfo {author} {\bibfnamefont {M.~C.}\
  \bibnamefont {Mooij}}, \bibinfo {author} {\bibfnamefont {R.~G.}\ \bibnamefont
  {Timmermans}}, \bibinfo {author} {\bibfnamefont {A.}~\bibnamefont {Touwen}},
  \bibinfo {author} {\bibfnamefont {W.}~\bibnamefont {Ubachs}}, \bibinfo
  {author} {\bibfnamefont {L.}~\bibnamefont {Willmann}}, \bibinfo {author}
  {\bibfnamefont {Y.}~\bibnamefont {Yin}},\ and\ \bibinfo {author}
  {\bibfnamefont {A.}~\bibnamefont {Zapara}},\ }\bibfield  {title} {\bibinfo
  {title} {{High accuracy theoretical investigations of CaF, SrF, and BaF and
  implications for laser-cooling}},\ }\href {https://doi.org/10.1063/1.5098540}
  {\bibfield  {journal} {\bibinfo  {journal} {Journal of Chemical Physics}\
  }\textbf {\bibinfo {volume} {151}},\ \bibinfo {pages} {034302} (\bibinfo
  {year} {2019})},\ \Eprint {https://arxiv.org/abs/1904.02516}
  {arXiv:1904.02516} \BibitemShut {NoStop}%
\bibitem [{\citenamefont {Courageux}\ \emph {et~al.}(2022)\citenamefont
  {Courageux}, \citenamefont {Cournol}, \citenamefont {Comparat}, \citenamefont
  {de~Lesegno},\ and\ \citenamefont {Lignier}}]{Courageux2022}%
  \BibitemOpen
  \bibfield  {author} {\bibinfo {author} {\bibfnamefont {T.}~\bibnamefont
  {Courageux}}, \bibinfo {author} {\bibfnamefont {A.}~\bibnamefont {Cournol}},
  \bibinfo {author} {\bibfnamefont {D.}~\bibnamefont {Comparat}}, \bibinfo
  {author} {\bibfnamefont {B.~V.}\ \bibnamefont {de~Lesegno}},\ and\ \bibinfo
  {author} {\bibfnamefont {H.}~\bibnamefont {Lignier}},\ }\bibfield  {title}
  {\bibinfo {title} {Efficient rotational cooling of a cold beam of barium
  monofluoride},\ }\href {https://doi.org/10.1088/1367-2630/ac511a} {\bibfield
  {journal} {\bibinfo  {journal} {New Journal of Physics}\ }\textbf {\bibinfo
  {volume} {24}},\ \bibinfo {pages} {025007} (\bibinfo {year}
  {2022})}\BibitemShut {NoStop}%
\bibitem [{\citenamefont {Denis}\ \emph {et~al.}(2022)\citenamefont {Denis},
  \citenamefont {Haase}, \citenamefont {Mooij}, \citenamefont {Chamorro},
  \citenamefont {Aggarwal}, \citenamefont {Bethlem}, \citenamefont
  {Boeschoten}, \citenamefont {Borschevsky}, \citenamefont {Esajas},
  \citenamefont {Hao}, \citenamefont {Hoekstra}, \citenamefont {van Hofslot},
  \citenamefont {Marshall}, \citenamefont {Meijknecht}, \citenamefont
  {Timmermans}, \citenamefont {Touwen}, \citenamefont {Ubachs}, \citenamefont
  {Willmann},\ and\ \citenamefont {Yin}}]{Denis2022}%
  \BibitemOpen
  \bibfield  {author} {\bibinfo {author} {\bibfnamefont {M.}~\bibnamefont
  {Denis}}, \bibinfo {author} {\bibfnamefont {P.~A.~B.}\ \bibnamefont {Haase}},
  \bibinfo {author} {\bibfnamefont {M.~C.}\ \bibnamefont {Mooij}}, \bibinfo
  {author} {\bibfnamefont {Y.}~\bibnamefont {Chamorro}}, \bibinfo {author}
  {\bibfnamefont {P.}~\bibnamefont {Aggarwal}}, \bibinfo {author}
  {\bibfnamefont {H.~L.}\ \bibnamefont {Bethlem}}, \bibinfo {author}
  {\bibfnamefont {A.}~\bibnamefont {Boeschoten}}, \bibinfo {author}
  {\bibfnamefont {A.}~\bibnamefont {Borschevsky}}, \bibinfo {author}
  {\bibfnamefont {K.}~\bibnamefont {Esajas}}, \bibinfo {author} {\bibfnamefont
  {Y.}~\bibnamefont {Hao}}, \bibinfo {author} {\bibfnamefont {S.}~\bibnamefont
  {Hoekstra}}, \bibinfo {author} {\bibfnamefont {J.~W.~F.}\ \bibnamefont {van
  Hofslot}}, \bibinfo {author} {\bibfnamefont {V.~R.}\ \bibnamefont
  {Marshall}}, \bibinfo {author} {\bibfnamefont {T.~B.}\ \bibnamefont
  {Meijknecht}}, \bibinfo {author} {\bibfnamefont {R.~G.~E.}\ \bibnamefont
  {Timmermans}}, \bibinfo {author} {\bibfnamefont {A.}~\bibnamefont {Touwen}},
  \bibinfo {author} {\bibfnamefont {W.}~\bibnamefont {Ubachs}}, \bibinfo
  {author} {\bibfnamefont {L.}~\bibnamefont {Willmann}},\ and\ \bibinfo
  {author} {\bibfnamefont {Y.}~\bibnamefont {Yin}} (\bibinfo {collaboration}
  {NL-$e\text{EDM}$ Collaboration}),\ }\bibfield  {title} {\bibinfo {title}
  {Benchmarking of the fock-space coupled-cluster method and uncertainty
  estimation: Magnetic hyperfine interaction in the excited state of {BaF}},\
  }\href {https://doi.org/10.1103/PhysRevA.105.052811} {\bibfield  {journal}
  {\bibinfo  {journal} {Phys. Rev. A}\ }\textbf {\bibinfo {volume} {105}},\
  \bibinfo {pages} {052811} (\bibinfo {year} {2022})}\BibitemShut {NoStop}%
\bibitem [{\citenamefont {Li}\ \emph {et~al.}(2023)\citenamefont {Li},
  \citenamefont {Ramachandran}, \citenamefont {Anderson},\ and\ \citenamefont
  {Vutha}}]{Li2023}%
  \BibitemOpen
  \bibfield  {author} {\bibinfo {author} {\bibfnamefont {S.~J.}\ \bibnamefont
  {Li}}, \bibinfo {author} {\bibfnamefont {H.~D.}\ \bibnamefont
  {Ramachandran}}, \bibinfo {author} {\bibfnamefont {R.}~\bibnamefont
  {Anderson}},\ and\ \bibinfo {author} {\bibfnamefont {A.~C.}\ \bibnamefont
  {Vutha}},\ }\bibfield  {title} {\bibinfo {title} {Optical control of baf
  molecules trapped in neon ice},\ }\href
  {https://doi.org/10.1088/1367-2630/ace9f3} {\bibfield  {journal} {\bibinfo
  {journal} {New Journal of Physics}\ }\textbf {\bibinfo {volume} {25}},\
  \bibinfo {pages} {082001} (\bibinfo {year} {2023})}\BibitemShut {NoStop}%
\bibitem [{\citenamefont {Shaw}\ \emph {et~al.}(2021)\citenamefont {Shaw},
  \citenamefont {Schnaubelt},\ and\ \citenamefont {McCarron}}]{Shaw2021}%
  \BibitemOpen
  \bibfield  {author} {\bibinfo {author} {\bibfnamefont {J.~C.}\ \bibnamefont
  {Shaw}}, \bibinfo {author} {\bibfnamefont {J.~C.}\ \bibnamefont
  {Schnaubelt}},\ and\ \bibinfo {author} {\bibfnamefont {D.~J.}\ \bibnamefont
  {McCarron}},\ }\bibfield  {title} {\bibinfo {title} {Resonance {Raman}
  optical cycling for high-fidelity fluorescence detection of molecules},\
  }\href {https://doi.org/10.1103/PhysRevResearch.3.L042041} {\bibfield
  {journal} {\bibinfo  {journal} {Phys. Rev. Research}\ }\textbf {\bibinfo
  {volume} {3}},\ \bibinfo {pages} {L042041} (\bibinfo {year}
  {2021})}\BibitemShut {NoStop}%
\bibitem [{\citenamefont {Zeng}\ \emph {et~al.}(2023)\citenamefont {Zeng},
  \citenamefont {Jadbabaie}, \citenamefont {Patel}, \citenamefont {Yu},
  \citenamefont {Steimle},\ and\ \citenamefont {Hutzler}}]{Zeng2023}%
  \BibitemOpen
  \bibfield  {author} {\bibinfo {author} {\bibfnamefont {Y.}~\bibnamefont
  {Zeng}}, \bibinfo {author} {\bibfnamefont {A.}~\bibnamefont {Jadbabaie}},
  \bibinfo {author} {\bibfnamefont {A.~N.}\ \bibnamefont {Patel}}, \bibinfo
  {author} {\bibfnamefont {P.}~\bibnamefont {Yu}}, \bibinfo {author}
  {\bibfnamefont {T.~C.}\ \bibnamefont {Steimle}},\ and\ \bibinfo {author}
  {\bibfnamefont {N.~R.}\ \bibnamefont {Hutzler}},\ }\bibfield  {title}
  {\bibinfo {title} {Optical cycling in polyatomic molecules with complex
  hyperfine structure},\ }\href {https://doi.org/10.1103/PhysRevA.108.012813}
  {\bibfield  {journal} {\bibinfo  {journal} {Phys. Rev. A}\ }\textbf {\bibinfo
  {volume} {108}},\ \bibinfo {pages} {012813} (\bibinfo {year}
  {2023})}\BibitemShut {NoStop}%
\bibitem [{\citenamefont {Fitch}\ \emph {et~al.}(2021)\citenamefont {Fitch},
  \citenamefont {Lim}, \citenamefont {Hinds}, \citenamefont {Sauer},\ and\
  \citenamefont {Tarbutt}}]{Fitch2020}%
  \BibitemOpen
  \bibfield  {author} {\bibinfo {author} {\bibfnamefont {N.~J.}\ \bibnamefont
  {Fitch}}, \bibinfo {author} {\bibfnamefont {J.}~\bibnamefont {Lim}}, \bibinfo
  {author} {\bibfnamefont {E.~A.}\ \bibnamefont {Hinds}}, \bibinfo {author}
  {\bibfnamefont {B.~E.}\ \bibnamefont {Sauer}},\ and\ \bibinfo {author}
  {\bibfnamefont {M.~R.}\ \bibnamefont {Tarbutt}},\ }\bibfield  {title}
  {\bibinfo {title} {{Methods for measuring the electron's electric dipole
  moment using ultracold YbF molecules}},\ }\href
  {https://doi.org/10.1088/2058-9565/abc931} {\bibfield  {journal} {\bibinfo
  {journal} {Quantum Science and Technology}\ }\textbf {\bibinfo {volume}
  {6}},\ \bibinfo {pages} {014006} (\bibinfo {year} {2021})},\ \Eprint
  {https://arxiv.org/abs/2009.00346} {arXiv:2009.00346} \BibitemShut {NoStop}%
\bibitem [{\citenamefont {Alauze}\ \emph {et~al.}(2021)\citenamefont {Alauze},
  \citenamefont {Lim}, \citenamefont {Trigatzis}, \citenamefont {Swarbrick},
  \citenamefont {Collings}, \citenamefont {Fitch}, \citenamefont {Sauer},\ and\
  \citenamefont {Tarbutt}}]{Alauze2021}%
  \BibitemOpen
  \bibfield  {author} {\bibinfo {author} {\bibfnamefont {X.}~\bibnamefont
  {Alauze}}, \bibinfo {author} {\bibfnamefont {J.}~\bibnamefont {Lim}},
  \bibinfo {author} {\bibfnamefont {M.~A.}\ \bibnamefont {Trigatzis}}, \bibinfo
  {author} {\bibfnamefont {S.}~\bibnamefont {Swarbrick}}, \bibinfo {author}
  {\bibfnamefont {F.~J.}\ \bibnamefont {Collings}}, \bibinfo {author}
  {\bibfnamefont {N.~J.}\ \bibnamefont {Fitch}}, \bibinfo {author}
  {\bibfnamefont {B.~E.}\ \bibnamefont {Sauer}},\ and\ \bibinfo {author}
  {\bibfnamefont {M.~R.}\ \bibnamefont {Tarbutt}},\ }\bibfield  {title}
  {\bibinfo {title} {An ultracold molecular beam for testing fundamental
  physics},\ }\href {https://doi.org/10.1088/2058-9565/ac107e} {\bibfield
  {journal} {\bibinfo  {journal} {Quantum Science and Technology}\ }\textbf
  {\bibinfo {volume} {6}},\ \bibinfo {pages} {044005} (\bibinfo {year}
  {2021})}\BibitemShut {NoStop}%
\bibitem [{\citenamefont {Augenbraun}\ \emph {et~al.}(2023)\citenamefont
  {Augenbraun}, \citenamefont {Anderegg}, \citenamefont {Hallas}, \citenamefont
  {Lasner}, \citenamefont {Vilas},\ and\ \citenamefont
  {Doyle}}]{Augenbraun2023}%
  \BibitemOpen
  \bibfield  {author} {\bibinfo {author} {\bibfnamefont {B.~L.}\ \bibnamefont
  {Augenbraun}}, \bibinfo {author} {\bibfnamefont {L.}~\bibnamefont
  {Anderegg}}, \bibinfo {author} {\bibfnamefont {C.}~\bibnamefont {Hallas}},
  \bibinfo {author} {\bibfnamefont {Z.~D.}\ \bibnamefont {Lasner}}, \bibinfo
  {author} {\bibfnamefont {N.~B.}\ \bibnamefont {Vilas}},\ and\ \bibinfo
  {author} {\bibfnamefont {J.~M.}\ \bibnamefont {Doyle}},\ }\bibfield  {title}
  {\bibinfo {title} {Chapter two - direct laser cooling of polyatomic
  molecules},\ }in\ \href
  {https://doi.org/https://doi.org/10.1016/bs.aamop.2023.04.005} {\emph
  {\bibinfo {booktitle} {Advances in Atomic, Molecular, and Optical
  Physics}}},\ \bibinfo {series} {Advances In Atomic, Molecular, and Optical
  Physics}, Vol.~\bibinfo {volume} {72},\ \bibinfo {editor} {edited by\
  \bibinfo {editor} {\bibfnamefont {L.~F.}\ \bibnamefont {DiMauro}}, \bibinfo
  {editor} {\bibfnamefont {H.}~\bibnamefont {Perrin}},\ and\ \bibinfo {editor}
  {\bibfnamefont {S.~F.}\ \bibnamefont {Yelin}}}\ (\bibinfo  {publisher}
  {Academic Press},\ \bibinfo {year} {2023})\ pp.\ \bibinfo {pages}
  {89--182}\BibitemShut {NoStop}%
\bibitem [{\citenamefont {Kogel}\ \emph {et~al.}(2021)\citenamefont {Kogel},
  \citenamefont {Rockenhäuser}, \citenamefont {Albrecht},\ and\ \citenamefont
  {Langen}}]{Kogel2021}%
  \BibitemOpen
  \bibfield  {author} {\bibinfo {author} {\bibfnamefont {F.}~\bibnamefont
  {Kogel}}, \bibinfo {author} {\bibfnamefont {M.}~\bibnamefont
  {Rockenhäuser}}, \bibinfo {author} {\bibfnamefont {R.}~\bibnamefont
  {Albrecht}},\ and\ \bibinfo {author} {\bibfnamefont {T.}~\bibnamefont
  {Langen}},\ }\bibfield  {title} {\bibinfo {title} {A laser cooling scheme for
  precision measurements using fermionic barium monofluoride (137ba19f)
  molecules},\ }\href {https://doi.org/10.1088/1367-2630/ac1df2} {\bibfield
  {journal} {\bibinfo  {journal} {New Journal of Physics}\ }\textbf {\bibinfo
  {volume} {23}},\ \bibinfo {pages} {095003} (\bibinfo {year}
  {2021})}\BibitemShut {NoStop}%
\bibitem [{\citenamefont {{Garcia Ruiz \textit{et
  al.}}}(2020)}]{GarciaRuiz2020}%
  \BibitemOpen
  \bibfield  {author} {\bibinfo {author} {\bibfnamefont {R.}~\bibnamefont
  {{Garcia Ruiz \textit{et al.}}}},\ }\bibfield  {title} {\bibinfo {title}
  {{Spectroscopy of short-lived radioactive molecules}},\ }\href
  {https://doi.org/10.1038/s41586-020-2299-4} {\bibfield  {journal} {\bibinfo
  {journal} {Nature}\ }\textbf {\bibinfo {volume} {581}},\ \bibinfo {pages}
  {396} (\bibinfo {year} {2020})}\BibitemShut {NoStop}%
\bibitem [{\citenamefont {Udrescu}\ \emph {et~al.}(2024)\citenamefont
  {Udrescu}, \citenamefont {Wilkins}, \citenamefont {Breier}, \citenamefont
  {Athanasakis-Kaklamanakis}, \citenamefont {Garcia~Ruiz}, \citenamefont {Au},
  \citenamefont {Belo{\v s}evi{\'c}}, \citenamefont {Berger}, \citenamefont
  {Bissell}, \citenamefont {Binnersley}, \citenamefont {Brinson}, \citenamefont
  {Chrysalidis}, \citenamefont {Cocolios}, \citenamefont {de~Groote},
  \citenamefont {Dorne}, \citenamefont {Flanagan}, \citenamefont {Franchoo},
  \citenamefont {Gaul}, \citenamefont {Geldhof}, \citenamefont {Giesen},
  \citenamefont {Hanstorp}, \citenamefont {Heinke}, \citenamefont
  {Koszor{\'u}s}, \citenamefont {Kujanp{\"a}{\"a}}, \citenamefont {Lalanne},
  \citenamefont {Neyens}, \citenamefont {Nichols}, \citenamefont {Perrett},
  \citenamefont {Reilly}, \citenamefont {Rothe}, \citenamefont {van~den Borne},
  \citenamefont {Vernon}, \citenamefont {Wang}, \citenamefont {Wessolek},
  \citenamefont {Yang},\ and\ \citenamefont {Z{\"u}lch}}]{Udrescu2024}%
  \BibitemOpen
  \bibfield  {author} {\bibinfo {author} {\bibfnamefont {S.~M.}\ \bibnamefont
  {Udrescu}}, \bibinfo {author} {\bibfnamefont {S.~G.}\ \bibnamefont
  {Wilkins}}, \bibinfo {author} {\bibfnamefont {A.~A.}\ \bibnamefont {Breier}},
  \bibinfo {author} {\bibfnamefont {M.}~\bibnamefont
  {Athanasakis-Kaklamanakis}}, \bibinfo {author} {\bibfnamefont {R.~F.}\
  \bibnamefont {Garcia~Ruiz}}, \bibinfo {author} {\bibfnamefont
  {M.}~\bibnamefont {Au}}, \bibinfo {author} {\bibfnamefont {I.}~\bibnamefont
  {Belo{\v s}evi{\'c}}}, \bibinfo {author} {\bibfnamefont {R.}~\bibnamefont
  {Berger}}, \bibinfo {author} {\bibfnamefont {M.~L.}\ \bibnamefont {Bissell}},
  \bibinfo {author} {\bibfnamefont {C.~L.}\ \bibnamefont {Binnersley}},
  \bibinfo {author} {\bibfnamefont {A.~J.}\ \bibnamefont {Brinson}}, \bibinfo
  {author} {\bibfnamefont {K.}~\bibnamefont {Chrysalidis}}, \bibinfo {author}
  {\bibfnamefont {T.~E.}\ \bibnamefont {Cocolios}}, \bibinfo {author}
  {\bibfnamefont {R.~P.}\ \bibnamefont {de~Groote}}, \bibinfo {author}
  {\bibfnamefont {A.}~\bibnamefont {Dorne}}, \bibinfo {author} {\bibfnamefont
  {K.~T.}\ \bibnamefont {Flanagan}}, \bibinfo {author} {\bibfnamefont
  {S.}~\bibnamefont {Franchoo}}, \bibinfo {author} {\bibfnamefont
  {K.}~\bibnamefont {Gaul}}, \bibinfo {author} {\bibfnamefont {S.}~\bibnamefont
  {Geldhof}}, \bibinfo {author} {\bibfnamefont {T.~F.}\ \bibnamefont {Giesen}},
  \bibinfo {author} {\bibfnamefont {D.}~\bibnamefont {Hanstorp}}, \bibinfo
  {author} {\bibfnamefont {R.}~\bibnamefont {Heinke}}, \bibinfo {author}
  {\bibfnamefont {{\'A}.}~\bibnamefont {Koszor{\'u}s}}, \bibinfo {author}
  {\bibfnamefont {S.}~\bibnamefont {Kujanp{\"a}{\"a}}}, \bibinfo {author}
  {\bibfnamefont {L.}~\bibnamefont {Lalanne}}, \bibinfo {author} {\bibfnamefont
  {G.}~\bibnamefont {Neyens}}, \bibinfo {author} {\bibfnamefont
  {M.}~\bibnamefont {Nichols}}, \bibinfo {author} {\bibfnamefont {H.~A.}\
  \bibnamefont {Perrett}}, \bibinfo {author} {\bibfnamefont {J.~R.}\
  \bibnamefont {Reilly}}, \bibinfo {author} {\bibfnamefont {S.}~\bibnamefont
  {Rothe}}, \bibinfo {author} {\bibfnamefont {B.}~\bibnamefont {van~den
  Borne}}, \bibinfo {author} {\bibfnamefont {A.~R.}\ \bibnamefont {Vernon}},
  \bibinfo {author} {\bibfnamefont {Q.}~\bibnamefont {Wang}}, \bibinfo {author}
  {\bibfnamefont {J.}~\bibnamefont {Wessolek}}, \bibinfo {author}
  {\bibfnamefont {X.~F.}\ \bibnamefont {Yang}},\ and\ \bibinfo {author}
  {\bibfnamefont {C.}~\bibnamefont {Z{\"u}lch}},\ }\bibfield  {title} {\bibinfo
  {title} {Precision spectroscopy and laser-cooling scheme of a
  radium-containing molecule},\ }\href
  {https://doi.org/10.1038/s41567-023-02296-w} {\bibfield  {journal} {\bibinfo
  {journal} {Nature Physics}\ }\textbf {\bibinfo {volume} {20}},\ \bibinfo
  {pages} {202} (\bibinfo {year} {2024})}\BibitemShut {NoStop}%
\bibitem [{\citenamefont {Pultinevicius}\ \emph {et~al.}(2023)\citenamefont
  {Pultinevicius}, \citenamefont {Rockenhäuser}, \citenamefont {Kogel},
  \citenamefont {Groß}, \citenamefont {Garg}, \citenamefont {Prochnow},\ and\
  \citenamefont {Langen}}]{Pultinevicius2023}%
  \BibitemOpen
  \bibfield  {author} {\bibinfo {author} {\bibfnamefont {E.}~\bibnamefont
  {Pultinevicius}}, \bibinfo {author} {\bibfnamefont {M.}~\bibnamefont
  {Rockenhäuser}}, \bibinfo {author} {\bibfnamefont {F.}~\bibnamefont
  {Kogel}}, \bibinfo {author} {\bibfnamefont {P.}~\bibnamefont {Groß}},
  \bibinfo {author} {\bibfnamefont {T.}~\bibnamefont {Garg}}, \bibinfo {author}
  {\bibfnamefont {O.~E.}\ \bibnamefont {Prochnow}},\ and\ \bibinfo {author}
  {\bibfnamefont {T.}~\bibnamefont {Langen}},\ }\bibfield  {title} {\bibinfo
  {title} {{A scalable scanning transfer cavity laser stabilization scheme
  based on the Red Pitaya STEMlab platform}},\ }\href
  {https://doi.org/10.1063/5.0169021} {\bibfield  {journal} {\bibinfo
  {journal} {Review of Scientific Instruments}\ }\textbf {\bibinfo {volume}
  {94}},\ \bibinfo {pages} {103004} (\bibinfo {year} {2023})}\BibitemShut
  {NoStop}%
\end{thebibliography}%
\end{document}